\documentclass[preprint,pra,showpacs]{revtex4-1}
\usepackage{amssymb}
\usepackage{amsmath}
\usepackage{eurosym}
\usepackage{graphicx}
\usepackage{dcolumn}
\usepackage{bm}

\begin{document}

\title{Interactions of spatial solitons with fused couplers}
\author{Alon Harel and Boris A. Malomed}
\address{Department of Physical Electronics, School of Electrical
Engineering, Faculty of Engineering, Tel Aviv University, Tel Aviv
69978, Israel}

\begin{abstract}
We study dynamical and stationary states of solitons in dual-core waveguides
which are locally coupled (\textit{fused}) at one or several short segments.
The model applies to planar optical waveguides, and to BEC in dual traps.
Collisions of an incident soliton with single and double locally-fused
couplers are investigated by means of systematic simulations and several
analytical methods (quasi-linear, fast-soliton, and adiabatic
approximations). Excitation dynamics of a soliton trapped by a local coupler
is studied by means of the variational approximation, and verified by
simulations. Shuttle motion of a soliton trapped in a cavity between two
local couplers, and in a finite array of couplers, is studied too.
\end{abstract}

 \maketitle

%\noindent OCIS numbers: 190.6135; 230.4320; 230.7390; 020.1475 % \pacs{}

\section{Introduction}

One of basic types of optical waveguides is represented by dual-core
couplers, in which parallel guiding cores interact via evanescent
fields \cite{Huang}. If the intrinsic nonlinearity in the cores is
strong enough, the power exchange between them is effected by the
intensity of the guided signals \cite{Jensen}, which is a basis for
the design of all-optical switching devices
\cite{switch}-\cite{review} and other applications, such as
nonlinear amplifiers \cite{amplifier}, stabilization of WDM
transmission schemes \cite{WDM}, and logic gates \cite{logic}.
%and bistable transmission \cite{Leon}
In addition to the simplest dual-core system, realizations of
nonlinear couplers have been proposed in many other settings,
including the use of orthogonal polarizations of light
\cite{Trillo}, semiconductor waveguides \cite{semi}, twin-core Bragg
gratings \cite{Bragg}, \cite{Sukhorukov}, systems with saturable
\cite{satur}, quadratic \cite{chi2,Shapira}, and
cubic-quintic (CQ) \cite{Albuch} nonlinearities, plasmonic media \cite%
{plasma}, dual-core traps for matter waves in Bose-Einstein condensates
(BEC) \cite{BEC1,Marek}, parallel arrays of discrete waveguides \cite%
{Herring}, nonlocal intra-core nonlinearity \cite{Nonlocal},
couplers for spatiotemporal ``light bullets" in dual-core planar
waveguides \cite{Dror}, and $\mathcal{PT}$-symmetric nonlinear couplers \cite%
{PT1}.

A fundamental property of nonlinear couplers with symmetric cores is the
\textit{symmetry-breaking bifurcation} (SBB), which destabilizes obvious
symmetric modes and gives rise to asymmetric ones. The SBB was analyzed, at
first, for spatially uniform states \cite{Snyder}, and then for solitons in
twin-core regular waveguides \cite{Wabnitz}-\cite{Kip} and Bragg gratings
\cite{Bragg} with the Kerr (cubic) nonlinearity (see also an early review
\cite{Wabnitz2}, and a more advanced one \cite{Progress}). The SBB\ analysis
was then extended to solitons in couplers with the quadratic \cite{chi2} and
CQ \cite{Albuch} nonlinearities.

The Kerr nonlinearity in the dual-core system gives rise to the \textit{%
subcritical} SBB for solitons, with originally unstable branches of emerging
asymmetric modes going backward (to weaker nonlinearity) and then turning
forward \cite{bifurcations}. The asymmetric modes retrieve the stability at
the turning points. On the other hand, the \textit{supercritical} SBB gives
rise to stable branches of asymmetric solitons going in the forward
direction. The SBB of the latter type\ occurs in twin-core Bragg grating
\cite{Bragg}, and in the case of the quadratic nonlinearity \cite{chi2}. The
system with the CQ nonlinearity gives rise to a \textit{bifurcation loop},
whose shape may be concave or convex \cite{Albuch}.

In addition to the numerical analysis of soliton modes in nonlinear
dual-core systems, the SBB point was found in an exact analytical form for
the system with the cubic nonlinearity \cite{Wabnitz}, and the emerging
asymmetric modes were studied by means of the variational approximation
(VA), see original works \cite{Laval,Muschall,satur,chi2,Bragg,Dror} and
review \cite{Progress}.

In addition to the studies of solitons in uniform dual-core systems,
analysis was also developed for \textit{fused couplers}, in which the two
cores are joined in a narrow segment \cite{X}. In the simplest
approximation, the corresponding dependence of the coupling strength on
transverse coordinate $x$ may be approximated by the delta-function, $\delta
(x)$. In previous works, interactions of solitons with such a locally fused
segment were studied in the temporal domain, \textit{viz}., for bright \cite%
{X,India} and dark \cite{dark} solitons in dual-core optical fibers and
fiber lasers \cite{laser}. In that case, the coupling affects the solitons
only over a short interval of their evolution.

Another possibility, which was discussed earlier for various types of
optical media \cite{Akhm,Raymond}, is to consider the \emph{spatial-domain}
dynamics of optical signals carried by dual-core planar waveguides with
short fused segments. Such structures can be molded using polymer materials
\cite{polymer1}, or built into photonic crystals by means of techniques
proposed in Refs. \cite{PhotCryst}. Similar structures are available in
plasmonics \cite{plasmonics}. An alternative is to use \textit{virtual}
dual-core guiding patterns, written in photorefractive media by means of
strong pump beams, on top of which probe beams propagate \cite{Neshev}. In
such settings, one can consider both stationary spatial solitons trapped by
the fused segment of the coupler, and scattering of incident spatial
solitons on one or several fused segments. This is the subject of the
present work.

The paper is organized as follows. The model is formulated in Section II,
where we also indicate that it applies as well to matter-wave couplers for
trapped BEC. Collisions of the incident soliton with a local coupler are
studied in Section III, by means of systematic simulations and three
analytical methods, which are relevant in different parametric regions
(quasi-linear, fast-soliton, and adiabatic approximations). Stationary and
excited states of a soliton trapped by the local coupler are considered in
Section IV, using the VA combined with a numerical approach. Shuttle
oscillations of a soliton trapped between two separated couplers, and in a
finite array of couplers, are studied in Section V. The paper is concluded
by Section VI.

\section{The model}

\begin{figure}[bh]
\includegraphics[width=3.5in]{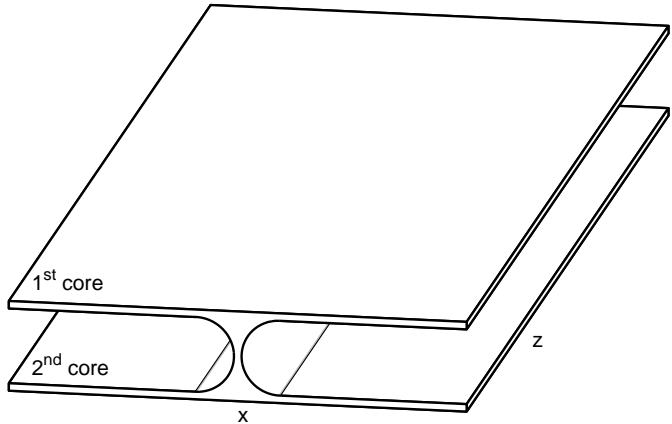}
\caption{A schematic structure of the fused planar coupler.}
\label{fig1}
\end{figure}

The propagation of electromagnetic waves with amplitudes $u\left( x,z\right)
$ and $v\left( x,z\right) $ along direction $z$ in the dual-core planar
nonlinear waveguide, fused along a narrow stripe around $x=0$, which is
approximated by the $\delta $-function (see a schematic shape of the fused
coupler in Fig. \ref{fig1}), obeys coupled nonlinear Schr\"{o}dinger (NLS)
equations, that can be derived in the scaled form, following the lines of
Ref. \cite{Boardman}:

\begin{align}
i\frac{\partial u}{\partial z}& =-\tfrac{1}{2}\frac{\partial ^{2}u}{\partial
x^{2}}-|u|^{2}u-\delta \left( x\right) v,  \label{nlse1} \\
i\frac{\partial v}{\partial z}& =-\tfrac{1}{2}\frac{\partial ^{2}v}{\partial
x^{2}}-|v|^{2}v-\delta \left( x\right) u.  \label{nlse2}
\end{align}%
Here the second derivatives represent the paraxial diffraction in the
transverse direction ($x$), and the equations are scaled so as to make the
coefficients in front of the nonlinear and coupling terms equal to $1$.
These equations can be derived from Lagrangian $L=\int_{-\infty }^{+\infty }%
\mathcal{L}dx,$ with density%
\begin{gather}
\mathcal{L}=\tfrac{i}{2}\left( u^{\ast }u_{z}-u_{z}^{\ast }u\right) -\tfrac{1%
}{2}|u_{x}|^{2}+\tfrac{1}{2}|u|^{4}  \notag \\
+\tfrac{i}{2}\left( v^{\ast }v_{z}-v_{z}^{\ast }v\right) -\tfrac{1}{2}%
|v_{x}|^{2}+\tfrac{1}{2}|v|^{4}+\delta \left( x\right) \left( u^{\ast
}v+v^{\ast }u\right) .  \label{L-density}
\end{gather}

To relate the scaled form of the model to physical units, one can follow the
standard derivation procedure, starting from the wave equation for the
electromagnetic fields in the dual-core waveguide \cite%
{Boardman,Huang,Agrawal}. Straightforward analysis yields the following
relation between length $l$ of the fused segment, which is approximated by
the $\delta $-functions in Eqs. (\ref{nlse1}) and (\ref{nlse2}), the
coupling length of the fused waveguide, $Z_{\mathrm{coupl}}$, wavelength $%
\lambda $, and characteristic scale $X_{0}$ used for the rescaling (it
implies that the characteristic spatial width of solitons in physical units
is $\sim X_{0}$):%
\begin{equation}
l\sim \lambda Z_{\mathrm{coupl}}/\left( 2\pi X_{0}\right) .  \label{L}
\end{equation}%
For relevant values $Z_{\mathrm{coupl}}\sim 1$ mm \cite{Boardman,Agrawal}
and the typical width of the spatial soliton $X_{0}\sim 50$ $\mathrm{\mu }$%
m, Eq. (\ref{L}) demonstrates that the typical length of the fused
segment is $l\sim 3\lambda $, which is quite realistic in terms of
the experimental fabrication of the couplers \cite{fabrication}.

Coupled equations (\ref{nlse1}) and (\ref{nlse2})\ were simulated using the
standard split-step Fourier-transform algorithm, with $\delta (x)$
approximated by a narrow Gaussian, $\hat{\delta}\left( x\right) $, which is
subject to the normalization condition,
%The resolution along the x-axis is in the order of the coupler's FWHM.
%, thus the actual coupling shape is less smooth than the one shown in Fig.~\ref{FigGaussianCoupling}.
$\int_{-\infty }^{+\infty }\hat{\delta}\left( x\right) dx=1$. The width of
the regularized $\delta $-function was chosen to be essentially smaller than
the width of the solitons considered below. We have checked that, under this
condition, results of the simulations practically do not depend on the
regularization width.

A pair of two locally fused couplers, separated by distance $D$, is
described by the NLS equations with the $\delta $-functions multiplied by $%
1/2$, hence the system with $D=0$ goes over back into Eqs. (\ref{nlse1}) and
(\ref{nlse2}):
\begin{eqnarray}
i\frac{\partial u}{\partial z} &=&-\tfrac{1}{2}\frac{\partial ^{2}u}{%
\partial x^{2}}-|u|^{2}u-\tfrac{1}{2}\left[ \delta \left( x-\tfrac{D}{2}%
\right) +\delta \left( x+\tfrac{D}{2}\right) \right] v,  \label{u-dual} \\
i\frac{\partial v}{\partial z} &=&-\tfrac{1}{2}\frac{\partial ^{2}v}{%
\partial x^{2}}-|v|^{2}v-\tfrac{1}{2}\left[ \delta \left( x-\tfrac{D}{2}%
\right) +\delta \left( x+\tfrac{D}{2}\right) \right] u.  \label{v-dual}
\end{eqnarray}%
Further, an array built of $2N+1$ couplers is described by the extension of
Eqs. (\ref{u-dual}), (\ref{v-dual}) with {}

\begin{equation}
\tfrac{1}{2}\left[ \delta \left( x-\tfrac{D}{2}\right) +\delta \left( x+%
\tfrac{D}{2}\right) \right] \rightarrow C\sum_{n=-N}^{N}\delta \left(
x-nD\right) ,  \label{C}
\end{equation}%
where $C$ is a coupling constant. For solitons whose width is much larger
than the array's spacing, $D$, the comb structure in Eq. (\ref{C}) may be
approximated by a uniform coupling constant, $\bar{C}\equiv C/D$, which acts
inside the corresponding box,
\begin{equation}
|x|~\leq X_{\mathrm{box}}\equiv ND.  \label{Xbox}
\end{equation}

It is relevant to mention that Eqs. (\ref{nlse1}), (\ref{nlse2}) and (\ref%
{u-dual}), (\ref{v-dual}), with $z$ replaced by time $t$, are also
meaningful as Gross-Pitaevskii equations (GPEs) for BEC loaded into
parallel cigar-shaped traps, i.e., matter-wave couplers
\cite{BEC1,Marek}. In that
case, the localized coupling may be induced by a transverse laser beam \cite%
{Hamburg}. Analyzing the derivation starting from the underlying
three-dimensional GPE \cite{GPE}, it is straightforward to arrive at a
relation between the length of the fused area ($l$), characteristic scale $%
X_{0}$ (which, as well as in the case of the optical system, determines the
spatial size of the corresponding matter-wave solitons), and radius $%
a_{\perp }$ of the transverse confinement of the BEC in the
quasi-one-dimensional trap, all taken in physical units:%
\begin{equation}
l\sim a_{\perp }^{2}/X_{0}~,  \label{l}
\end{equation}%
cf. Eq. (\ref{L}). For typical values $a_{\perp }\sim 3$ $\mathrm{\mu }$m
and $X_{0}\sim 10~\mathrm{\mu }$m \cite{Randy}, Eq. (\ref{l}) yields $l\sim
1~\mathrm{\mu }$m, which is a relevant estimate for the size of the area
induced by focused laser beams.

\section{Collisions of the soliton with the single coupler}

\subsection{Numerical results}

The collision of the incident soliton, launched in the \textit{straight} ($u$%
) core, with the local coupler was simulated using Eqs. (\ref{nlse1}), (\ref%
{nlse2}) with the regularized $\delta $-function and the following initial
conditions:

\begin{equation}
u\left( x\right) |_{z=0}=\eta ~\mathrm{sech}\left( \eta \left( x-\xi
_{0}\right) \right) \exp \left( iqx\right) ,v\left( x\right) |_{z=0}=0.
\label{initialv}
\end{equation}%
Here $\eta $ is the amplitude of the soliton, $q$ is its velocity [in fact,
the tilt of the spatial soliton in the $\left( x,z\right) $ plane], and $\xi
_{0}<0$ with large $\left\vert \xi _{0}\right\vert \gg 1/\eta $ is the
initial position. The total power of the incident soliton is $P\equiv
\int_{-\infty }^{+\infty }\left[ \left\vert u(x)\right\vert ^{2}+\left\vert
v(x)\right\vert ^{2}\right] dx=2\eta .$

\begin{figure}[tbp]
$%
\begin{array}{cc}
\includegraphics[width=3.5in]{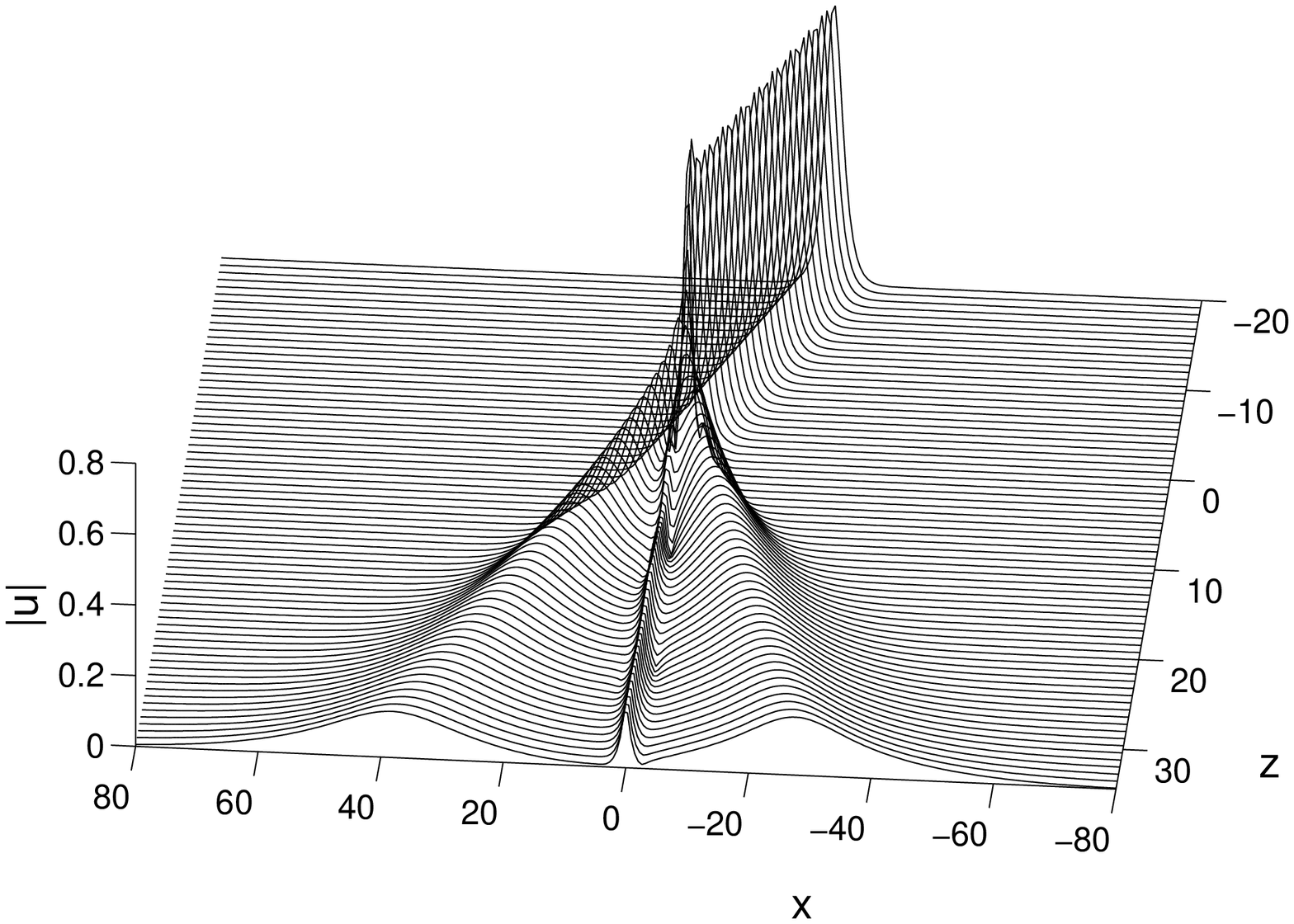} & %
\includegraphics[width=3.5in]{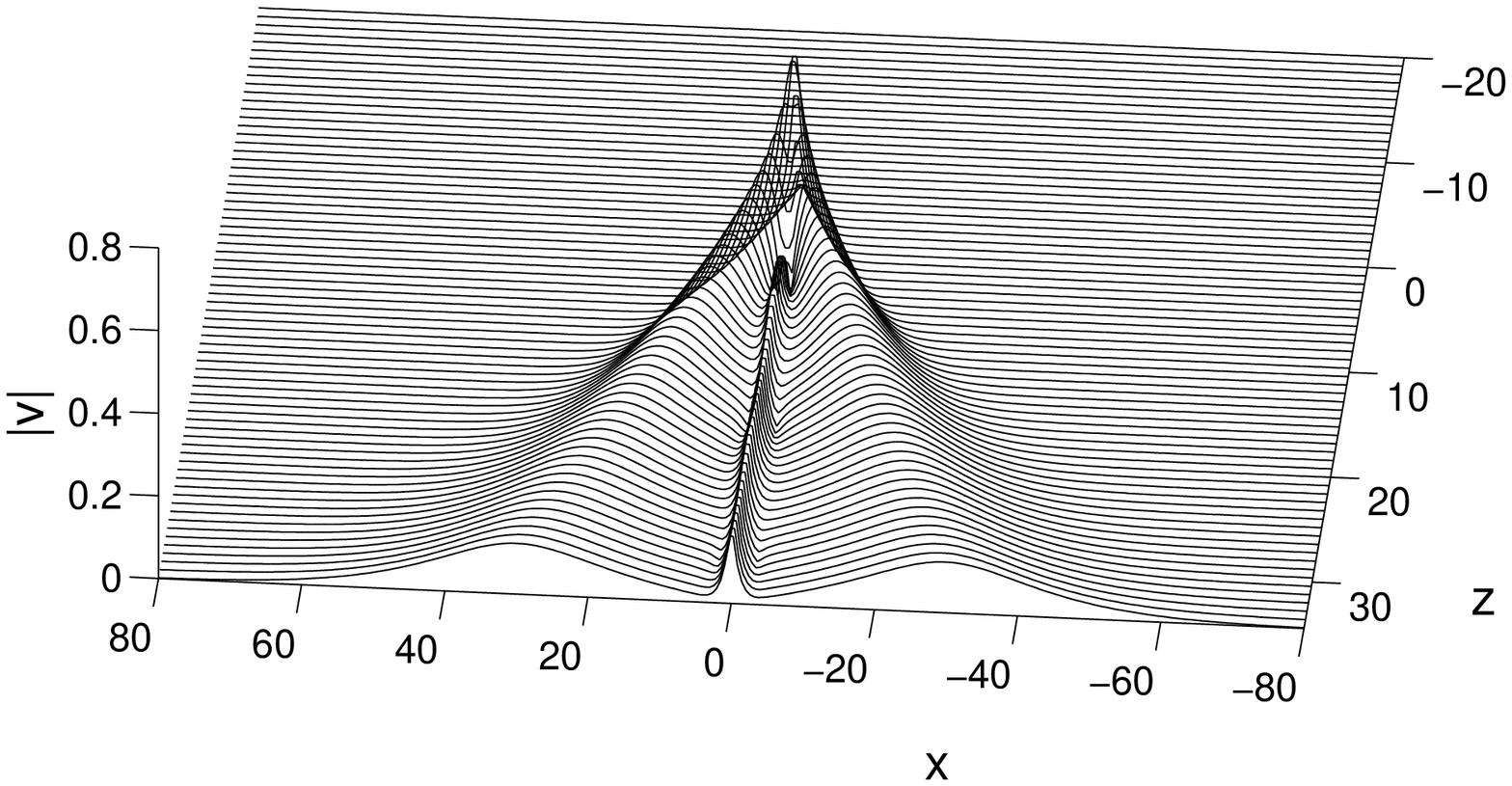} \\
(\mathrm{a}) & (\mathrm{b})%
\end{array}%
$%
\caption{Collision of an incident soliton with the fused coupler, for $%
\protect\eta =0.8$ and $q=1$ [see Eq. (\protect\ref{initialv})], in the
straight (a) and cross (b) cores.}
\label{FigSingleScatter}
\end{figure}

As shown in Fig. \ref{FigSingleScatter}, the soliton-coupler interaction
gives rise to five waves: transferred ($u_{T}$) and reflected ($u_{R}$) ones
in the straight core; their counterparts, $v_{T}$ and $v_{R}$, in the
\textit{cross} (second) core; and a trapped mode oscillating between the two
cores in a vicinity of $x=0$. Obviously, powers of these waves must obey the
conservation relation:

\begin{equation}
P\left( u_{T}\right) +P\left( u_{R}\right) +P\left( v_{T}\right) +P\left(
v_{R}\right) +P_{\mathrm{trap}}=2\eta .  \label{power}
\end{equation}

Results of systematic simulations are summarized in Fig. \ref%
{FigSingleCoupler}, which demonstrates the splitting of the total power of
the incident soliton between the five components. In particular, the power
shares carried by the transmitted and reflected waves in the cross core, $%
v_{T}$ and $v_{R}$, are equal, which is explained below. Panel (d) in this
figure demonstrates strong trapping for slowly moving ($q\lesssim 1$) heavy (%
$\eta \gtrsim 2.5$) incident solitons. At larger velocities, the interaction
of the soliton with the local coupler naturally weakens, and the power is
chiefly transferred in the straight core, as seen in panel (b). Lighter
incident solitons, with $\eta \lesssim 1$, do not generate the trapped mode,
being reflected in the straight core at $q\lesssim 0.5$, or transferred to
the cross core, where they are evenly split into the transmitted and
reflected components, at $q\gtrsim 0.5$. For heavy solitons, with $\eta
\gtrsim 3$, panels (b) and (d) exhibit a threshold value of the velocity, $%
q_{\mathrm{thr}}$, which is a sharp boundary between the trapping at $q<q_{%
\mathrm{thr}}$ and transmission at $q>q_{\mathrm{thr}}$.

\begin{figure}[tbp]
$%
\begin{array}{cc}
\includegraphics[width=3.5in]{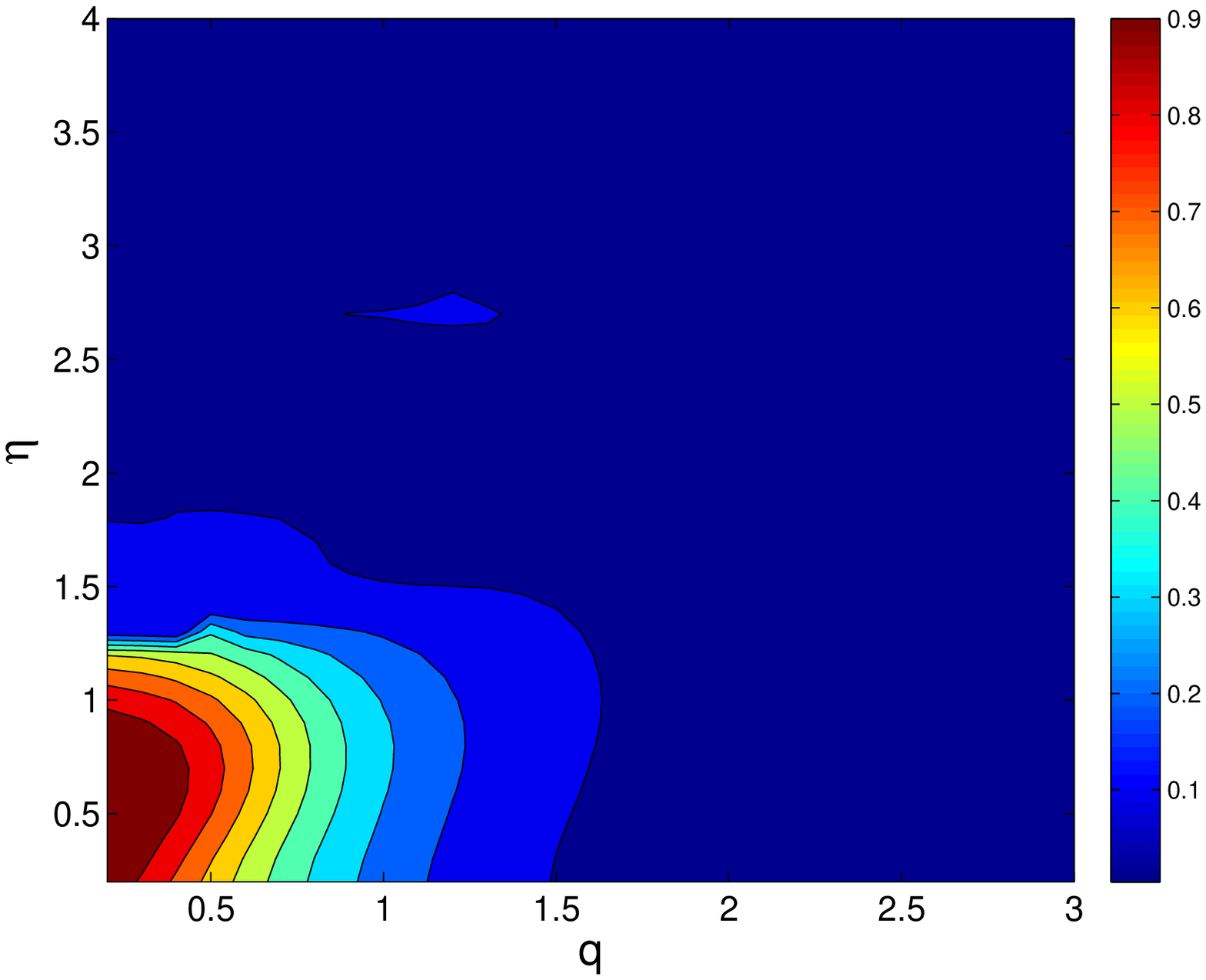} & %
\includegraphics[width=3.5in]{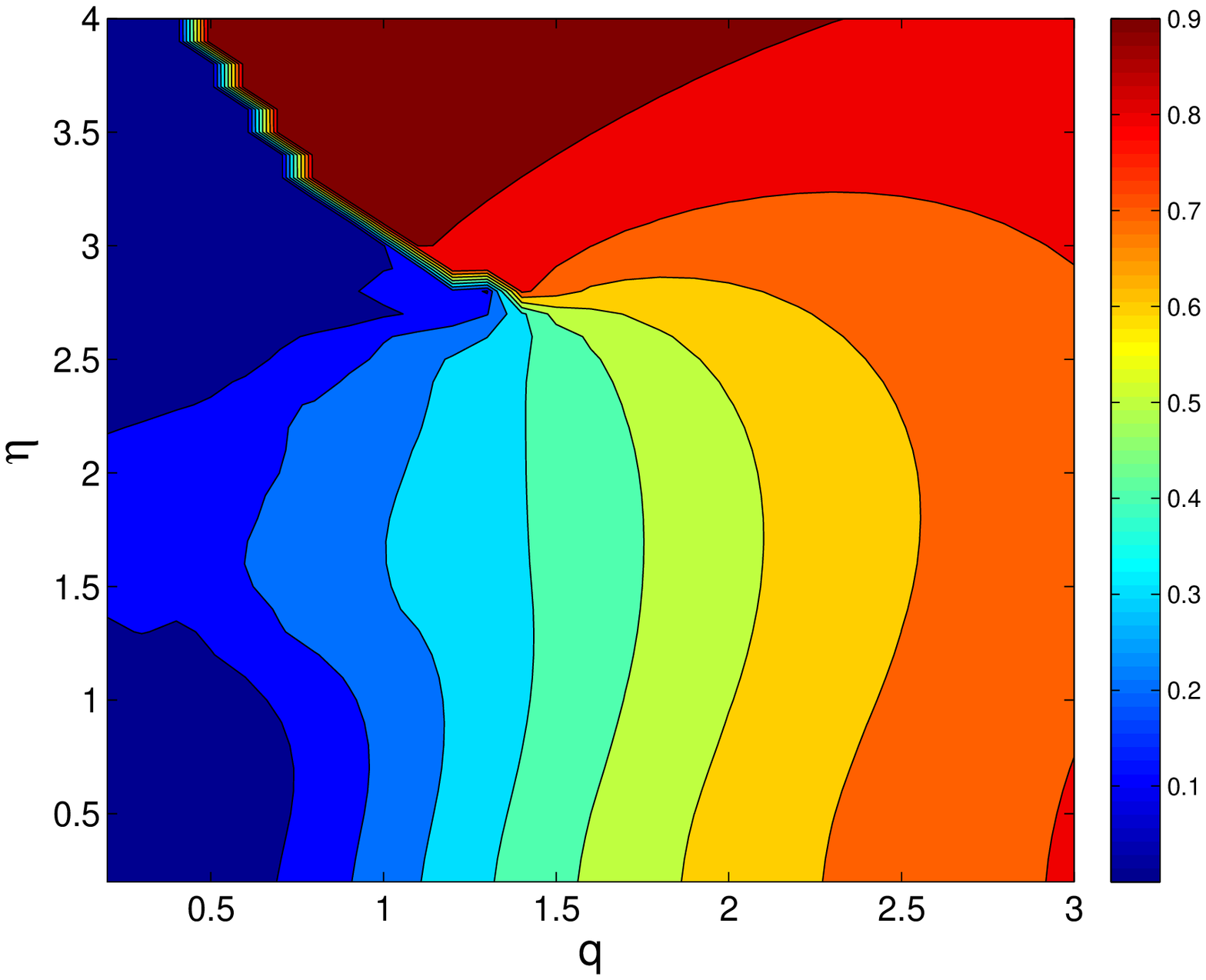} \\
(\mathrm{a}) & (\mathrm{b}) \\
\includegraphics[width=3.5in]{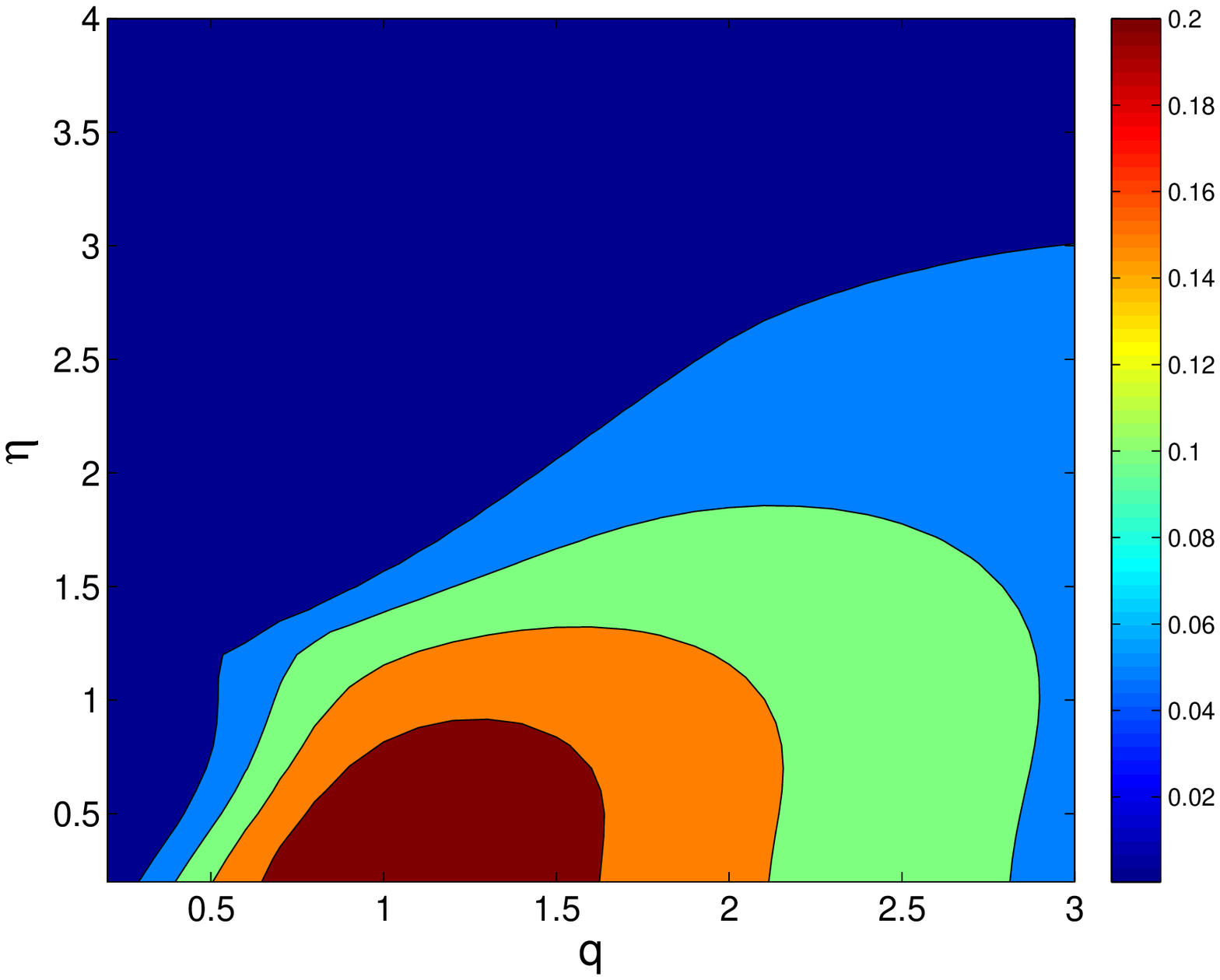} & %
\includegraphics[width=3.5in]{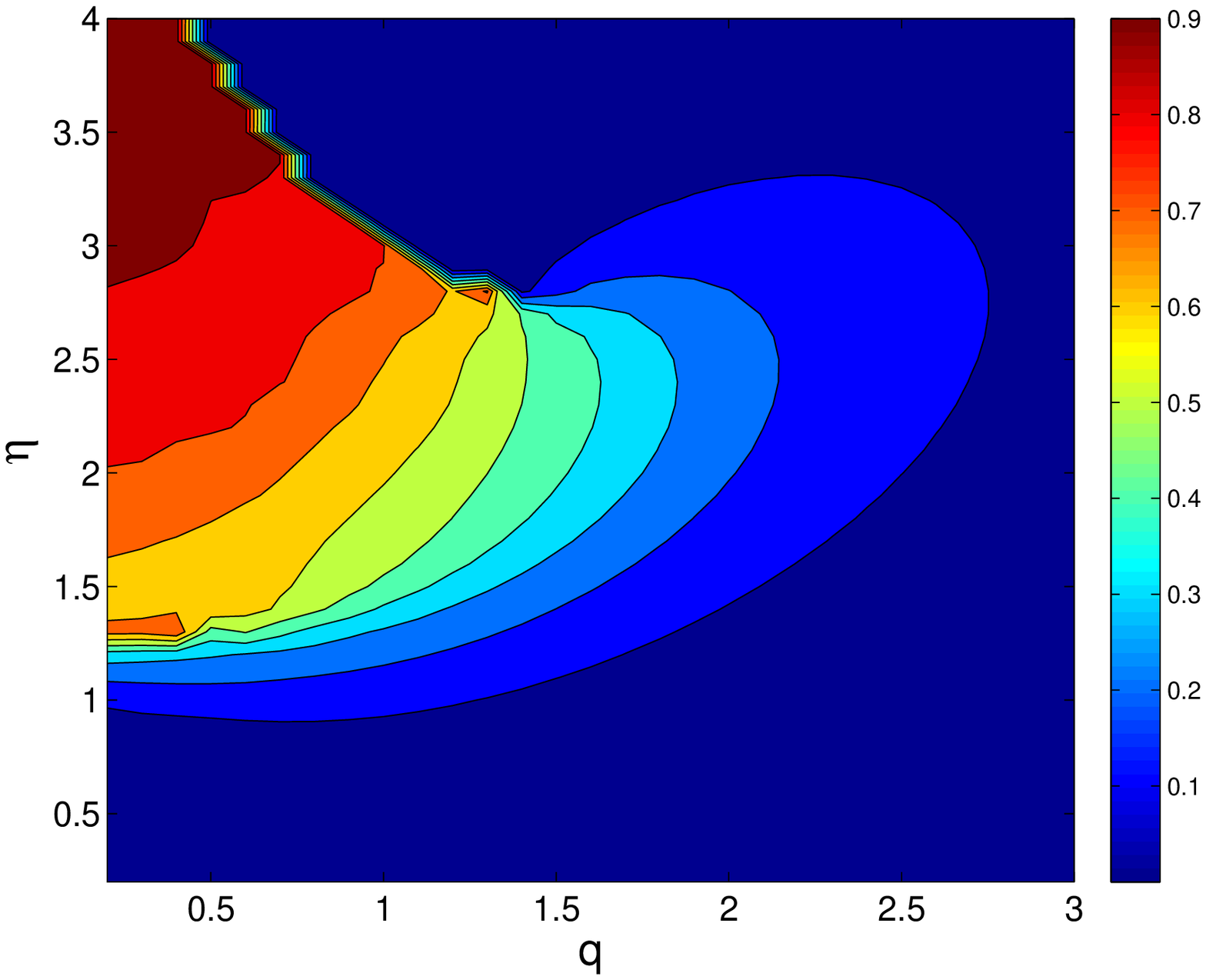} \\
(\mathrm{c}) & (\mathrm{d})%
\end{array}%
$%
\caption{(Color online) Fractions of the power of the incident soliton
reflected (a) and transmitted (b) in the straight core, scattered
(transmitted =\ reflected) in the cross core\ (c), and trapped by the fused
coupler (d), vs. the velocity and amplitude of the incident soliton, $q$ and
$\protect\eta $. The powers obey the balance condition (\protect\ref{power}%
). }
\label{FigSingleCoupler}
\end{figure}

\subsection{The quasi-linear approximation}

The nonlinearity in Eqs. (\ref{nlse1}), (~\ref{nlse2}) may be neglected in
comparison with the coupling terms if the potential energy of the nonlinear
self-interaction is much smaller than the inter-core coupling energy, which,
in the present notation, implies $\eta \ll 1$. Then, the substitution of $%
\{u\left( x,z\right) ,v\left( x,z\right) \}=\exp \left( ikz\right)
\{U(x),V(x)\}$, with propagation constant $k$, into the linearized version
of Eqs. (\ref{nlse1}), (~\ref{nlse2}) leads to the scattering problem based
on the stationary equations,

\begin{equation}
k\left\{ U,V\right\} =\tfrac{1}{2}\tfrac{d^{2}}{dx^{2}}\left\{ U,V\right\}
+\delta \left( x\right) \left\{ V,U\right\} .  \label{linear_pde_x2}
\end{equation}%
A solution to these equations, with wavenumber $q$ [cf. Eq. (\ref{initialv}%
)] and the respective value of the propagation constant,
\begin{equation}
k=-q^{2}/2,  \label{kq}
\end{equation}%
is sought for as
\begin{align}
U(x)& =\left\{
\begin{array}{lr}
U_{I}\exp \left( iqx\right) +U_{R}\exp \left( -iqx\right) , & x<0, \\
U_{T}\exp \left( iqx\right) , & x>0,%
\end{array}%
\right.  \label{linear_u_part} \\
V(x)& =\left\{
\begin{array}{lr}
V_{R}\exp \left( -iqx\right) , & x<0, \\
V_{T}\exp \left( iqx\right) , & x>0.%
\end{array}%
\right.  \label{linear_v_part}
\end{align}%
The amplitudes of the incident, transmitted , and reflected waves are
related by conditions of the continuity of $U(x)$ and $V(x)$ at $x=0$, $%
U_{I}+U_{R}=U_{T},~V_{R}=V_{T}$. With regard to this relation, the jump of
the first derivatives following from integration of Eq. (\ref{linear_pde_x2}%
) in an infinitesimal vicinity of $x=0$ yields another pair of equations for
the scattering amplitude, which amount to the form of $%
V_{T}=iq^{-1}U_{T}=-iqU_{R}$. %$ iqU_{R}=-iq^{-1}U_{T}=-V_{T}$.

A trapped mode is also generated by Eq. (\ref{linear_pde_x2}): $%
U=V=U_{0}\exp \left( -|x|\right) $, where $U_{0}$ is an arbitrary amplitude,
and the corresponding propagation constant is $k_{\mathrm{trap}}=1/2$.
Because it is separated from values (\ref{kq}), the trapped mode gives no
contribution to the linear scattering problem.

It is straightforward to find a solution to the continuity and jump
equations for the amplitudes:%
\begin{gather}
U_{R}=\left( q^{2}+1\right) ^{-1}U_{I},U_{T}=-\left( q^{2}+1\right)
^{-1}q^{2}U_{I},  \label{linear_frac_powers_first} \\
V_{R}=V_{T}=-i\left( q^{2}+1\right) ^{-1}qU_{I}.
\label{linear_frac_powers_last}
\end{gather}%
This solution satisfies the power-balance condition, cf. Eq. (\ref{power}),
and explains the above-mentioned symmetry between the transmitted and
reflected waves in the cross core. Further, in Fig.~\ref%
{FigLinearScatteringRatios} the comparison of the solution, given by Eqs.~(%
\ref{linear_frac_powers_first}) and~(\ref{linear_frac_powers_last}), to the
numerical results for the scattering of the soliton with a sufficiently
small amplitude, $\eta =0.1$, demonstrates a very close agreement between
the numerical and analytical results.
\begin{figure}[bh]
\includegraphics[width=3.5in]{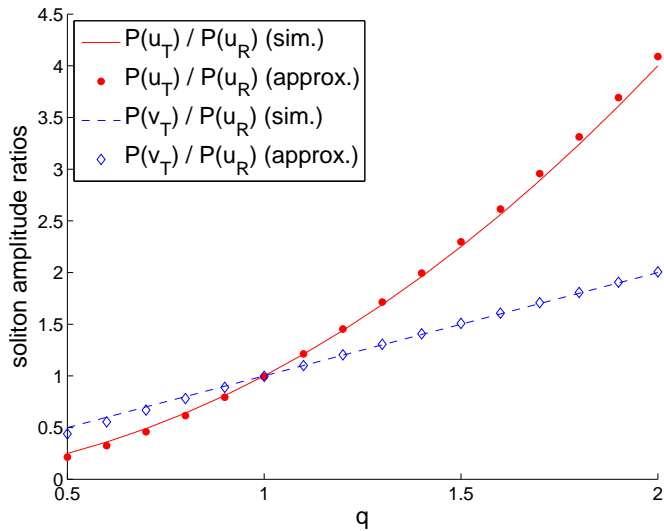}
\caption{(Color online) Comparison of the power ratios, given by the
analytical solution of the linear scattering problem, as per Eqs. (\protect
\ref{linear_frac_powers_first}) and (\protect\ref{linear_frac_powers_last})
(dashed curves), to the their counterparts produced by direct simulations
(chains of stars) for the small-amplitude soliton with $\protect\eta =0.1$.}
\label{FigLinearScatteringRatios}
\end{figure}

%\begin{figure}[tbp]
%$%
%\begin{array}{cc}
%\includegraphics[width=3.5in]{scattering_linear_1st_core.eps} & %
%\includegraphics[width=3.5in]{scattering_linear_2nd_core.eps} \\
%(\mathrm{a}) & (\mathrm{b})%
%\end{array}%
%$%
%\caption{Weak non-linearity scattering evolution (shown for $q=1,\protect%
%\eta =0.1$) , in the first (a) and second (b) cores. The reflected and
%transmitted solitons have the same amplitude, as expected from the above
%linear approximation.}
%\label{FigLinearScatteringEvolve}
%\end{figure}

\subsection{The interaction of a fast soliton with the local coupler}

Another analytical approximation applies to the collision of a fast soliton
with the local coupler, when the incident soliton itself remains almost
unaffected by the interaction, while generating weak transmitted and
reflected waves in the cross core. In the present notation, this case
amounts to taking $q\equiv Q\gg 1$ in initial condition (\ref{initialv}) [$q$
is replaced here by $Q$ to distinguish it from the wavenumber in dispersion
relation (\ref{kq})]. The corresponding incident NLS soliton is
\begin{equation}
u=\eta ~\mathrm{sech}\left( \eta \left( x-Qz\right) \right) e^{iQx}\exp
\left( \tfrac{1}{2}iz\left( \eta ^{2}-Q^{2}\right) \right) ,
\label{nls_soliton}
\end{equation}%
while Eq. (\ref{nlse2}) for the cross core may be linearized:
\begin{equation}
i\frac{\partial v}{\partial z}+\frac{1}{2}\frac{\partial ^{2}v}{\partial
x^{2}}=-\delta \left( x\right) u.  \label{nlse2fast}
\end{equation}

Thus, soliton (\ref{nls_soliton}), if substituted in Eq. (\ref{nlse2fast}),
gives rise to a localized source generating waves in the cross core. As
suggested by dispersion relation (\ref{kq}), the source generates two
distinct components of solutions, \textit{viz}., radiation, with $k<0$, and
trapped (non-propagating) modes with $k>0$. The so generated wave field can
be calculated by means of the Fourier transform. A final result is rather
cumbersome, as the emerging integrations cannot be performed analytically.
It takes a relatively simple form at $x=0$, where both radiation and trapped
fields have their maxima:
\begin{align}
v_{\mathrm{trap}}\left( x=0,z\right) & =-\frac{i}{2Q}\int_{0}^{\infty }\sqrt{%
\frac{2}{k}}\mathrm{sech}\left( \frac{\pi }{2\eta Q}\left( k+\frac{%
Q^{2}-\eta ^{2}}{2}\right) \right) e^{ikz}dk  \notag \\
& \approx \frac{\left[ 1-i~\mathrm{sgn}(z)\right] \sqrt{\pi }}{2Q\sqrt{|z|}}%
\mathrm{sech}\left( \frac{\pi \left( Q^{2}-\eta ^{2}\right) }{4\eta Q}%
\right) ,  \label{fast-trap} \\
v_{\mathrm{rad}}\left( x=0,z\right) & =\frac{i}{2Q}\int_{-\infty }^{0}\sqrt{%
\frac{2}{-k}}\mathrm{sech}\left( \frac{\pi }{2\eta Q}\left( k+\frac{%
Q^{2}-\eta ^{2}}{2}\right) \right) e^{ikz}dk  \notag \\
& \approx \frac{\left[ 1+i~\mathrm{sgn}(z)\right] \sqrt{\pi }}{2Q\sqrt{z}}%
\mathrm{sech}\left( \frac{\pi \left( Q^{2}-\eta ^{2}\right) }{4\eta Q}%
\right) .  \label{fast-rad}
\end{align}%
The asymptotic approximations in Eqs. (\ref{fast-trap}) and (\ref{fast-rad})
are valid for $|z|\gg \left( \eta Q\right) ^{-1}$. Because the present
situation is actually considered for large $Q$ and large $\eta $, the latter
approximations are always relevant, as a matter of fact.

This prediction is compared to results of direct simulations in Fig.~\ref%
{FigFastMotion}, where a close agreement is seen. At much larger values of $%
z $, the trapped radiation does not follow the asymptotic behavior predicted
by Eq. (\ref{fast-trap}), as a broad soliton eventually self-traps in the
cross core.

\begin{figure}[tbp]
\includegraphics[width=3.5in]{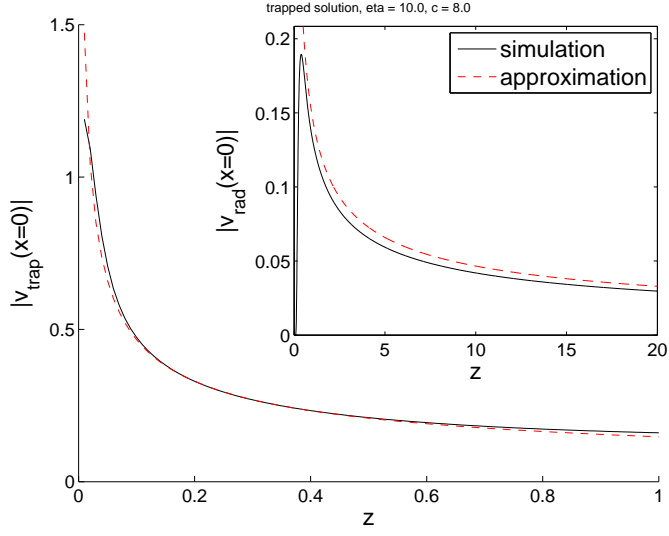}
\caption{(Color online) Comparison of the analytical results, given by Eqs. (%
\protect\ref{fast-trap}) and (\protect\ref{fast-rad}), to their numerical
counterparts for the absolute values of the trapped (main plot) and
radiation (inset) components of the field in the cross core at $x=0$,
produced by the passage of a fast heavy soliton (\protect\ref{nls_soliton}),
with $Q=8,\protect\eta =10$, through the local coupler. The comparison is
shown for $z>0$, as at $z<0$ (before the soliton passes the fused coupler)
it is difficult to separate the numerically generated field into the
radiation and trapped components.}
\label{FigFastMotion}
\end{figure}

\subsection{The adiabatic approximation}

In addition to the collision of fast incident solitons with the local
coupler, analytically tractable is also the case of a relatively slow heavy
soliton, with large $\eta $ and $q\ll \eta $ in Eq. (\ref{initialv}).
Accordingly, the soliton moving in the straight core is similar to the one
given by Eq. (\ref{nls_soliton}),
\begin{equation}
u=\eta ~\mathrm{sech}\left[ \eta \left( x-\xi \left( z\right) \right) \right]
\exp \left( i\xi ^{\prime }x\right) \exp \left( \tfrac{1}{2}iz\Lambda \right)
\label{adia_nls_soliton}
\end{equation}%
where $\xi ^{\prime }\equiv d\xi /dz$, and $\Lambda \equiv \left( \eta
^{2}-\xi ^{\prime }{}^{2}\right) $. Further, the field equation in the cross
core may be linearized as in Eq. (\ref{nlse2fast}) and, accordingly, the
solution component in this core, driven by the soliton's field (\ref%
{adia_nls_soliton}) via the local coupler, is easily found in the adiabatic
approximation, which treats $\xi (z)$ as a relatively slowly varying
function, and omits $\xi ^{\prime }{}^{2}$ in comparison with $\eta ^{2}$:

\begin{equation}
v\left( x,z\right) =\exp \left( i\eta ^{2}z/2\right) \mathrm{sech}\left(
\eta \xi (z)\right) \exp \left( -\eta |x|\right) .  \label{adia_ansatz_full}
\end{equation}

\begin{figure}[tbp]
\includegraphics[width=3.5in]{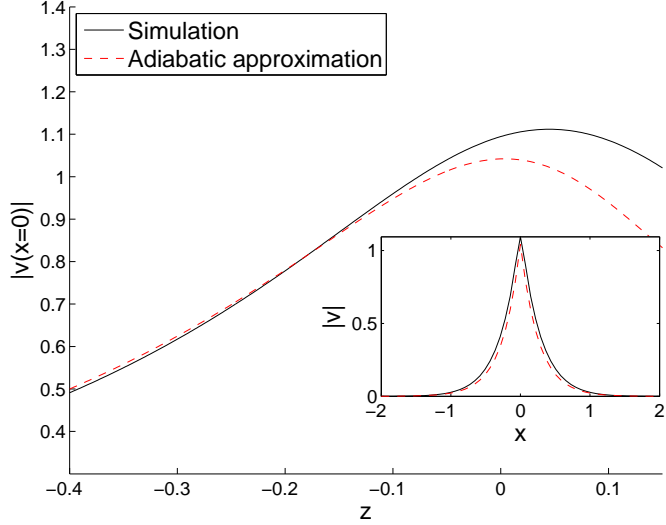}
\caption{(Color online) The comparison of the amplitude of the $v$-component
of the soliton, predicted Eq. (\protect\ref{adia_ansatz_full}) in the
adiabatic approximation, to its counterpart produced by direct simulations,
for amplitude $\protect\eta =4$ and initial velocity $q=0.456$.The inset
shows the comparison of full profiles of the $u$ and $v$ component at $z=0$,
when the soliton is passing the local coupler.}
\label{FigAdiaShape}
\end{figure}

The comparison with direct simulations, displayed in Fig. \ref{FigAdiaShape}%
, demonstrates that the amplitude and shape of expression (\ref%
{adia_ansatz_full}) are very close to those of the $v$-component of the
numerical solution up to the point of $\xi (z)=0$, where the incident
soliton hits the local coupler. Afterwards, a deviation emerges because the $%
v$-component does not completely vanish, as is formally predicted by Eq. (%
\ref{adia_ansatz_full}) at $\xi (z)\rightarrow \infty $.

Two terms in Lagrangian density (\ref{L-density}) contribute to the
interaction between the incident soliton and the local coupler. First, the
coupling term generates the attraction potential,
\begin{equation}
W_{\mathrm{coupl}}\equiv -\int_{-\infty }^{+\infty }\delta \left( x\right)
\left( u^{\ast }v+v^{\ast }u\right) dx=-2\eta ~\mathrm{sech}^{2}\left( \eta
\xi \right) ,  \label{attraction}
\end{equation}%
where the use was made of Eqs. (\ref{adia_nls_soliton}) and (\ref%
{adia_ansatz_full}). Second, the gradient energy of component (\ref%
{adia_ansatz_full}) gives rise to an effective repulsive potential, $W_{%
\mathrm{grad}}\equiv \tfrac{1}{2}\int_{-\infty }^{+\infty }|v_{x}|^{2}dx=%
\tfrac{1}{2}\eta ~\mathrm{sech}^{2}\left( \eta \xi \right) $. Thus, the net
potential of the interaction of the soliton with the local coupler is
attractive,
\begin{equation}
W_{\mathrm{tot}}(\xi )\equiv W_{\mathrm{coupl}}+W_{\mathrm{grad}}=-\tfrac{3}{%
2}\eta ~\mathrm{sech}^{2}\left( \eta \xi \right) .  \label{adia_full_inter}
\end{equation}%
The incident soliton accelerates under the action of this attraction force
and attains the largest velocity while passing the local coupler ($\xi =0$).
In the case of a small initial velocity, $q^{2}\ll 1$, see Eq. (\ref%
{initialv}), the velocity of the heavy soliton passing the coupler is thus
predicted by the energy-balance condition, $\left( M/2\right) \left( \xi
_{\max }^{\prime }\right) ^{2}=-W_{\mathrm{tot}}(\xi =0)$, where the
soliton's mass is $M=2\eta $. Thus, Eq. (\ref{adia_full_inter}) predicts $%
\xi _{\max }^{\prime }=\sqrt{3/2}$. The prediction is in agreement with
direct simulations, which demonstrate that this velocity ranges between $1.1$
and $1.2$.

In the adiabatic approximation, the reflection of the soliton from the local
coupler, which is represented by the potential well given by Eq. (\ref%
{adia_full_inter}), cannot be explained, although a reflection area is
clearly present in Fig. \ref{FigSingleCoupler} at $\eta \lesssim 1$, $%
q\lesssim 0.3$, and an example of periodic reflections is displayed below in
Fig. \ref{FigCavity} for $\eta =0.6$, $q=0.4$ (in those cases, the adiabatic
approximation is not relevant, as $\eta $ is too small). Actually, it is
known that \emph{reflection} of solitons from \emph{attractive} potentials
is possible beyond the framework of the adiabatic approximation, if the
interaction of the soliton with the trapped mode and radiation waves is
taken into regard \cite{Brand}.

\section{Stationary and excited modes trapped by the local coupler}

\subsection{The stationary modes}

Analytical trapped-mode (soliton) solutions to stationary equations (\ref%
{linear_pde_x2}) with $k>0$, which also include the nonlinear terms, $U^{3}$
and $V^{3}$, were found in Ref. \cite{Raymond}, in the form of
\begin{equation}
\{U\left( x\right) ,V\left( x\right) \}=\sqrt{2k}\mathrm{sech}\left[ \sqrt{2k%
}\left( |x|~+\{\xi ,\zeta \}\right) \right] .  \label{station_exact_form}
\end{equation}%
For the symmetric trapped \ mode, positive constant $\xi =\zeta $ is
determined by equation $\tanh \left( \sqrt{2k}\xi \right) =\left( 2k\right)
^{-1/2},$ hence the symmetric mode exists at $k>1/2$. On the other hand, as
said above, the SBB in the nonlinear coupler gives rise to asymmetric
solitons, which in the present case happens at $k_{\mathrm{SBB}}=3/2$, and
at $k>3/2$ there exist a pair of asymmetric solutions, which are mirror
images of each other. They are given by Eq. (\ref{station_exact_form}) with%
\begin{equation}
\mathrm{sech}^{2}\left( \sqrt{2k}\left\{ \xi ,\zeta \right\} \right) =\left(
4k\right) ^{-1}\left( 2k+1\pm \sqrt{4k^{2}-4k-3}\right) .  \label{tantisymm}
\end{equation}

\begin{figure}[tbp]
\includegraphics[width=3.5in]{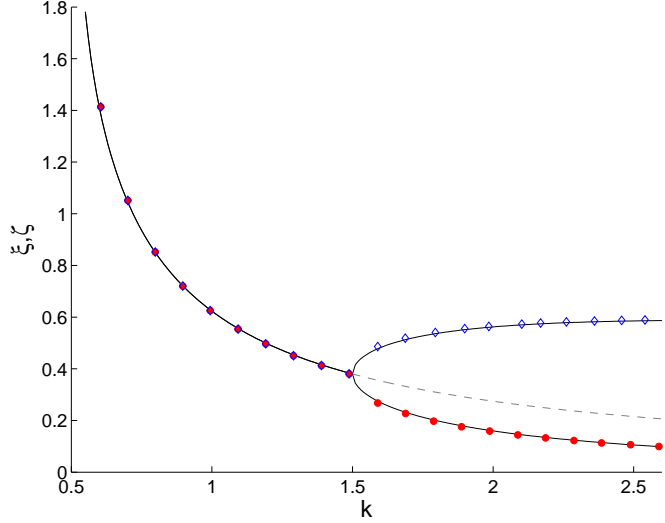}
\caption{(Color online) Exact stationary solutions for the symmetric and
asymmetric trapped modes, as given by Eqs. (\protect\ref{station_exact_form}%
)-(\protect\ref{tantisymm}), are represented by continuous curves.
Numerically found counterparts of these solutions are shown by chains of
symbols [the respective values of $\protect\xi $ and $\protect\zeta $ are
identified via $U(x=0)$ and $V(x=0)$ as per Eq. (\protect\ref%
{station_exact_form})]. The dashed line shows the unstable portion of the
symmetric-solution family. }
\label{Stationary}
\end{figure}

As shown in Ref. \cite{Raymond}, this SBB\ is of the supercritical type, but
these analytical predictions were not compared to numerical solutions
before, therefore we have done it here, see Fig. \ref{Stationary}. Further,
the simulations demonstrate that the symmetric solution at $1/2<k<3/2$, as
well as the asymmetric ones at $k>3/2$, are stable, while the symmetric
solution is unstable at $k>3/2$, as expected in the case of the
supercritical bifurcation \cite{bifurcations}.

\subsection{The variational approximation for excited modes}

In addition to stationary solutions (\ref{station_exact_form}) for trapped
modes, it is relevant to analyze excited states of such modes. This can be
done on the basis of the variational ansatz,
\begin{align}
u\left( x,z\right) & =\eta ~\mathrm{sech}\left[ \eta \left( x-\xi (z)\right) %
\right] \exp \left( i\phi (z)\right) ,  \label{vari_ansatz_u} \\
v\left( x,z\right) & =\theta ~\mathrm{sech}\left[ \theta \left( x-\zeta
(z)\right) \right] \exp \left( i\psi (z)\right)  \label{vari_ansatz_v}
\end{align}%
where $\eta $, $\theta $ and $\phi $, $\psi $ are (constant) amplitudes and
phases of the two components, while $\xi $ and $\zeta $ represent
excitations in the form of shifts of the components from the central
position, $x=0$.

Evolution equations for the so excited mode can be derived as Newton's
equations of motion for two interacting particles corresponding to the $u$
and $v$-components, which is equivalent to the VA \cite{Progress}. To this
end, we note that potential (\ref{attraction}) of the coupling between the
components yields, in the present case, $W_{\mathrm{coupl}}=-2\eta \theta
\mathrm{sech}\left( \theta \zeta \right) \mathrm{sech}\left( \eta \xi
\right) \cos \left( \phi -\psi \right) .$ The corresponding equations of
motion are $M_{\left\{ u,v\right\} }d^{2}\left\{ \xi ,\zeta \right\}
/dz^{2}=-\partial W_{\mathrm{coupl}}/\partial \left\{ \xi ,\zeta \right\} $,
where the above-mentioned effective soliton masses are $M_{\left\{
u,v\right\} }=2\left\{ \eta ,\theta \right\} $. Thus, the equations take the
form of
\begin{eqnarray}
\frac{d^{2}\xi }{dz^{2}} &=&-\eta \theta \frac{\sinh \left( \eta \xi \right)
\cos \left( \phi -\psi \right) }{\cosh ^{2}\left( \eta \xi \right) \cosh
\left( \theta \zeta \right) },  \label{vari_motion1} \\
\frac{d^{2}\zeta }{dz^{2}} &=&-\eta \theta \frac{\sinh \left( \theta \zeta
\right) \cos \left( \phi -\psi \right) }{\cosh ^{2}\left( \theta \zeta
\right) \cosh \left( \eta \xi \right) }.  \label{vari_motion2}
\end{eqnarray}%
Linearization of Eqs. (\ref{vari_motion1}) and (\ref{vari_motion2}) for
small oscillations of $\xi $ and $\zeta $ reduces to $d^{2}\xi /dz^{2}=-\eta
^{2}\theta \xi $ and $d^{2}\zeta /dz^{2}=-\eta \theta ^{2}\zeta $, hence
eigenfrequencies of the small positional oscillations are predicted to be
\begin{equation}
\Omega _{1}\approx \eta \sqrt{\theta },~\Omega _{2}=\theta \sqrt{\eta }.
\label{vari_approx_freq1}
\end{equation}

%
%Using the known effective mass of a soliton ($2\eta$ and $2\theta$, for $u$ and $v$ respectively),
%the motion equations can be written as:
%\begin{align}
%\frac{d^2 \xi}{d z^2} &= -\eta \theta \sech \left(\theta \zeta \right)
%\cos \left(\phi - \psi \right) \frac{\sinh \left( \eta \xi \right)}{\cosh^2 \left( \eta \xi \right)},
%\label{vari_motion1}  \\
%\frac{d^2 \zeta}{d z^2} &= -\eta \theta \sech \left(\eta \xi \right) \cos \left(\phi - \psi \right)
%\frac{\sinh \left( \theta \zeta \right)}{\cosh^2 \left( \theta \zeta \right)} \label{vari_motion2}
%\end{align}
%
%Eq.~\ref{vari_motion1} and ~\ref{vari_motion2} describe the attraction force applied
%to each soliton in its own core, and therefore will govern the solitons fluctuations along the x axis.
%Applying small perturbations approximation to these equations, one may obtain:
%\begin{align}
%\frac{d^2 \xi}{d z^2} & \approx -\eta^2 \theta \xi,  \label{vari_approx_motion1}  \\
%\frac{d^2 \xi}{d z^2} & \approx -\theta^2 \eta \zeta,  \label{vari_approx_motion2}
%\end{align}
%%
%yielding the perturbation frequencies:
%\begin{align}
%% Use the & sign before the = sign to align the two equations
%\Omega_1 & \approx \sqrt{\eta^2 \theta}, \label{vari_approx_freq1}  \\
%\Omega_2 & \approx \sqrt{\theta^2 \eta} \label{vari_approx_freq2}
%\end{align}

\begin{figure}[tbp]
\includegraphics[width=3.5in]{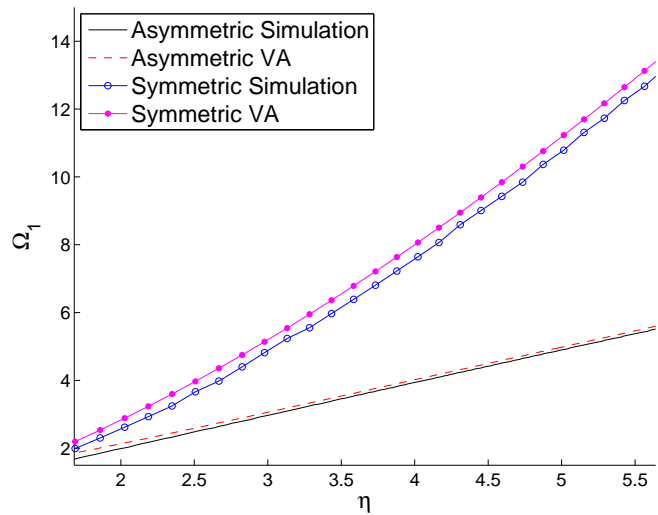}
\caption{(Color online) The comparison of the VA prediction for
eigenfrequencies of small positional oscillations of the components of the
symmetric ($\protect\eta =\protect\theta $, upper lines) and asymmetric ($%
\protect\eta \neq \protect\theta $, lower lines)\ trapped modes, as given by
Eq. (\protect\ref{vari_approx_freq1}), to results extracted from direct
simulations.}
\label{VariOmega1}
\end{figure}

These predictions were verified by simulations, as shown in Fig.~\ref%
{VariOmega1}. The simulations were run by applying a small kick to either
component of the symmetric or asymmetric trapped mode (\ref%
{station_exact_form}). In the asymmetric configuration, the positional
oscillations of the lighter component [the one with a smaller amplitude,
which corresponds to a smaller frequency in Eq. (\ref{vari_approx_freq1})]
exhibit strong attenuation, unlike robust oscillation of the stronger
component. Therefore, only one frequency is shown in Fig.~\ref{VariOmega1}
for the excited asymmetric mode.

%\begin{figure}[tbp]
%\includegraphics[width=3.5in]{w12_asym.eps}
%\caption{Core to core fluctuations in the asymmetric case ($\protect\eta\ne%
%\protect\theta$)}
%\label{FigVariBeat}
%\end{figure}

Small oscillations of the power between the two components (without position
shifts) were simulated too. It was found that the respective eigenfrequency
is very close to the obvious beating frequency, $\Omega _{12}=\left( \eta
^{2}-\theta ^{2}\right) /2$.

\section{Pairs and arrays of local couplers}

\subsection{The interaction of solitons with the double coupler}

\begin{figure}[tbp]
$%
\begin{array}{cc}
\includegraphics[width=3.5in]{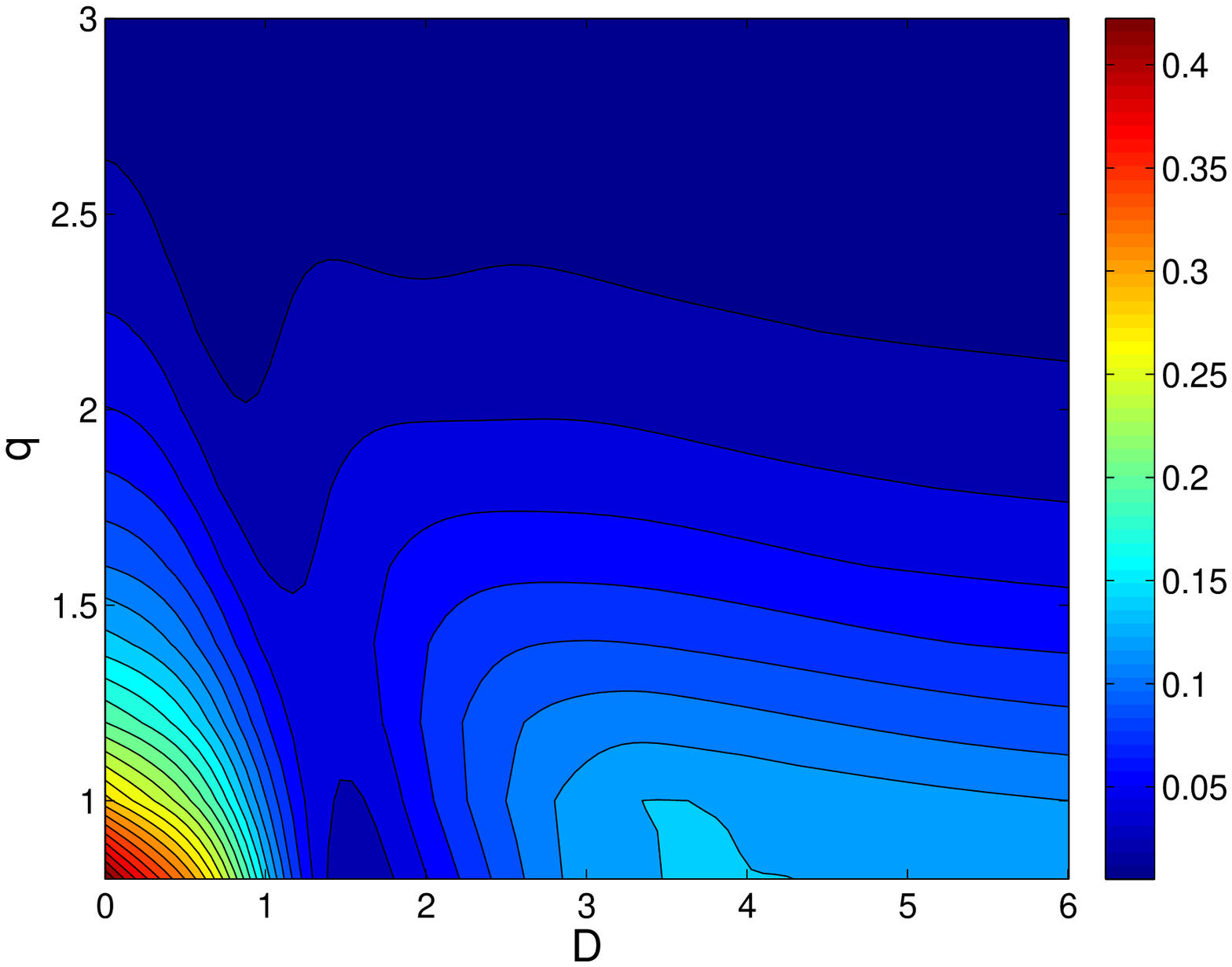} & %
\includegraphics[width=3.5in]{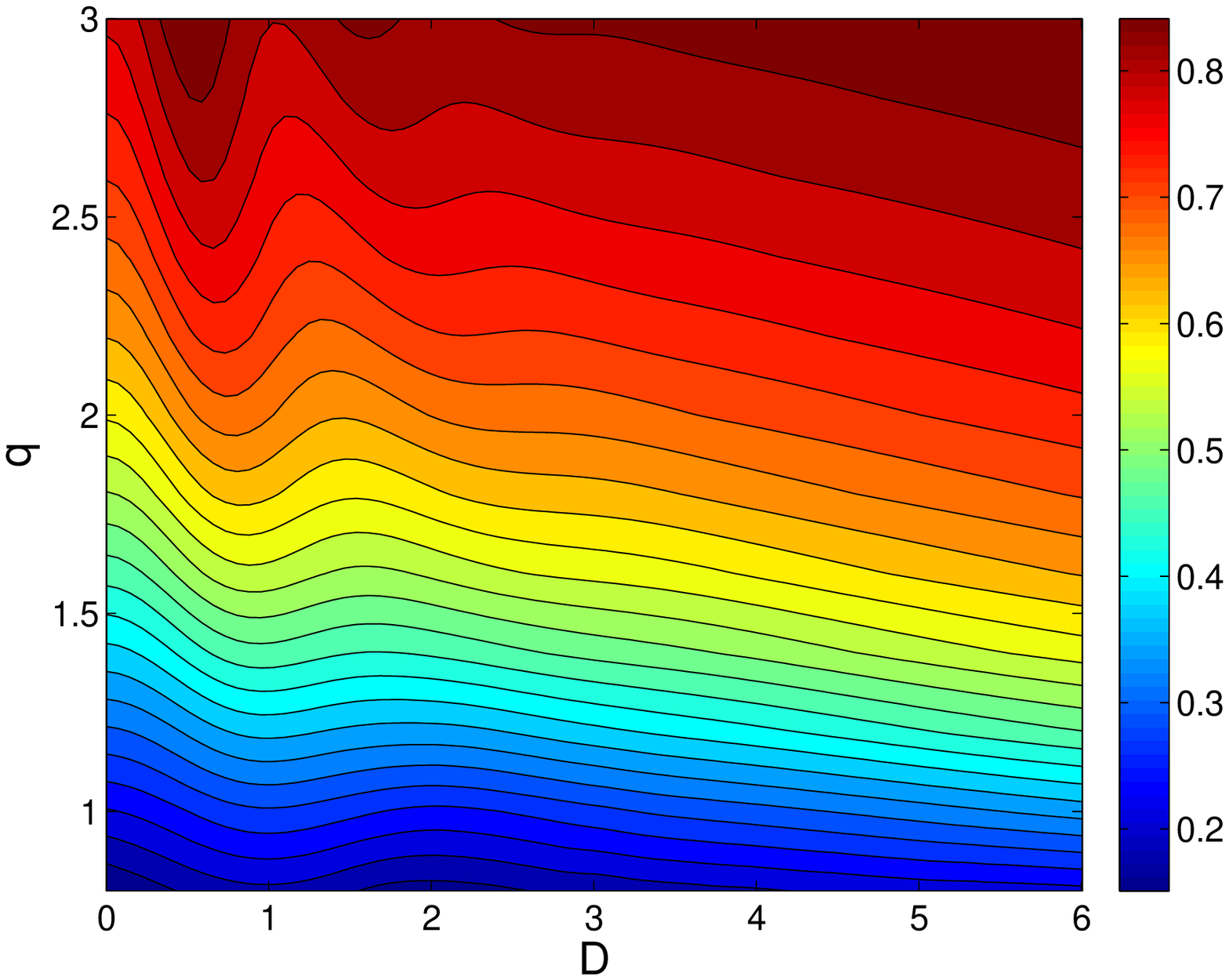} \\
(\mathrm{a}) & (\mathrm{b}) \\
\includegraphics[width=3.5in]{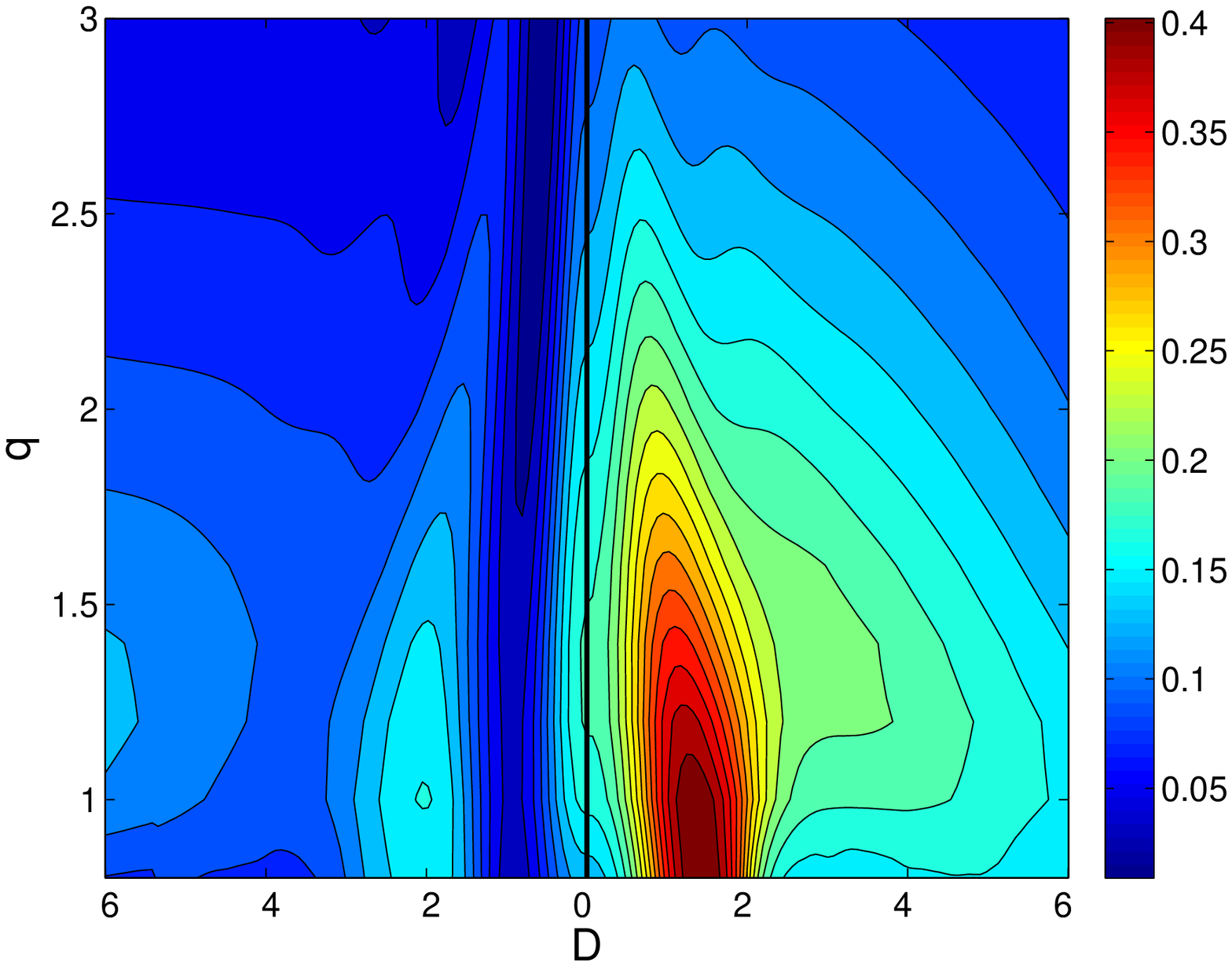} & %
\includegraphics[width=3.5in]{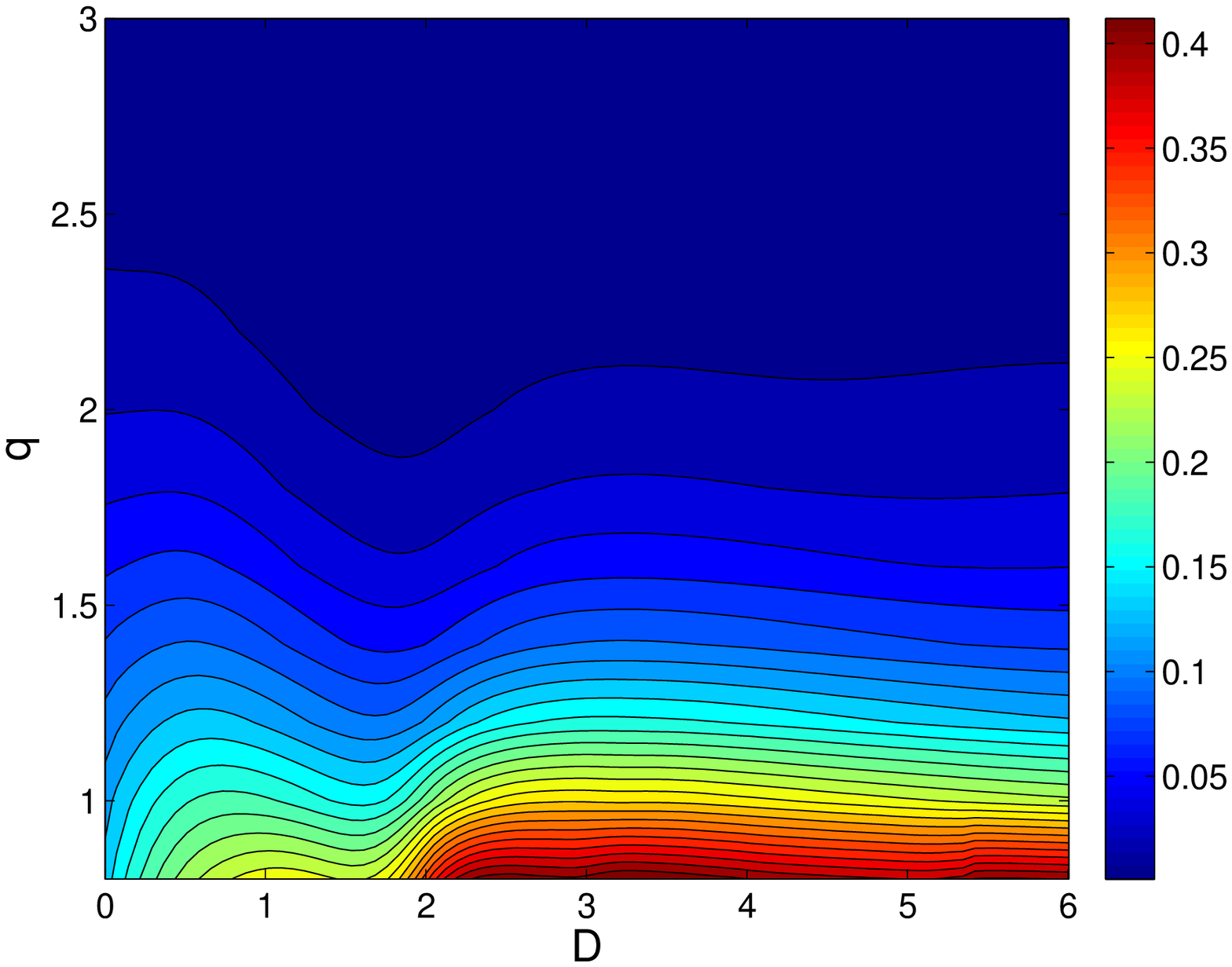} \\
(\mathrm{c}) & (\mathrm{d})%
\end{array}%
$%
\caption{(Color online) Splitting of the power of the incident soliton, with
amplitude $\protect\eta =1$ and velocity $q$, hitting the pair of local
couplers separated by distance $D$. (a,b) Shares of the power reflected and
transmitted, respectively, in the straight core. (c) The same for the cross
core. In (c), the plots for the reflected and transmitted shares are
butt-joined to corroborate the equality of these shares for the single
coupler ($D=0$). (d) The trapped-power share.}
\label{FigDualCoupler}
\end{figure}

Simulations of Eqs. (\ref{u-dual}) and (\ref{v-dual}) for the collision of
an incident soliton with the double locally-fused coupler were run using
initial condition (\ref{initialv}) with amplitude $\eta =1$, while both the
initial velocity, $q$, and distance $D$ between the two local couplers were
varied. Results of the simulations, summarized in Fig. \ref{FigDualCoupler},
show a conspicuous correlation between $u_{R}$ and $v_{T}$, i.e., the
reflected and transmitted waves in the straight and cross cores,
respectively. Similarly, a correlation is observed between the waves
transmitted in the straight core, $u_{T}$, and reflected in the cross core, $%
v_{R}$. It is worthy to note that the second coupler strongly affects the
scattering in the cross core, breaking the symmetry between the transmitted
and reflected waves in favor of the latter one, in comparison with the
single coupler, cf. Figs.~\ref{FigDualCoupler}(c) and \ref{FigSingleCoupler}%
(c). The share of the power of the incident soliton trapped by the pair of
local couplers, which is shown in Fig. \ref{FigDualCoupler}(d), was
evaluated long enough after the incident soliton hit the first local
coupler, \textit{viz}., at $z=175/q$ (recall $q$ is the incident soliton's
velocity). At relatively small velocities, the trapped share increases in
comparison with the single coupler ($D=0$).

% Use figure* when spanning a figure across the two columns

%\begin{figure}[tbp]
%\includegraphics[width=3.5in]{dual-core-contourf-trapped_no_title.eps}
%\caption{The same as in Fig. \protect\ref{FigDualCoupler}, but for the share
%of the power share trapped by the dual coupler.}
%\label{FigDualCouplerTrapped}
%\end{figure}

\begin{figure}[tbp]
$%
\begin{array}{cc}
\includegraphics[width=3in]{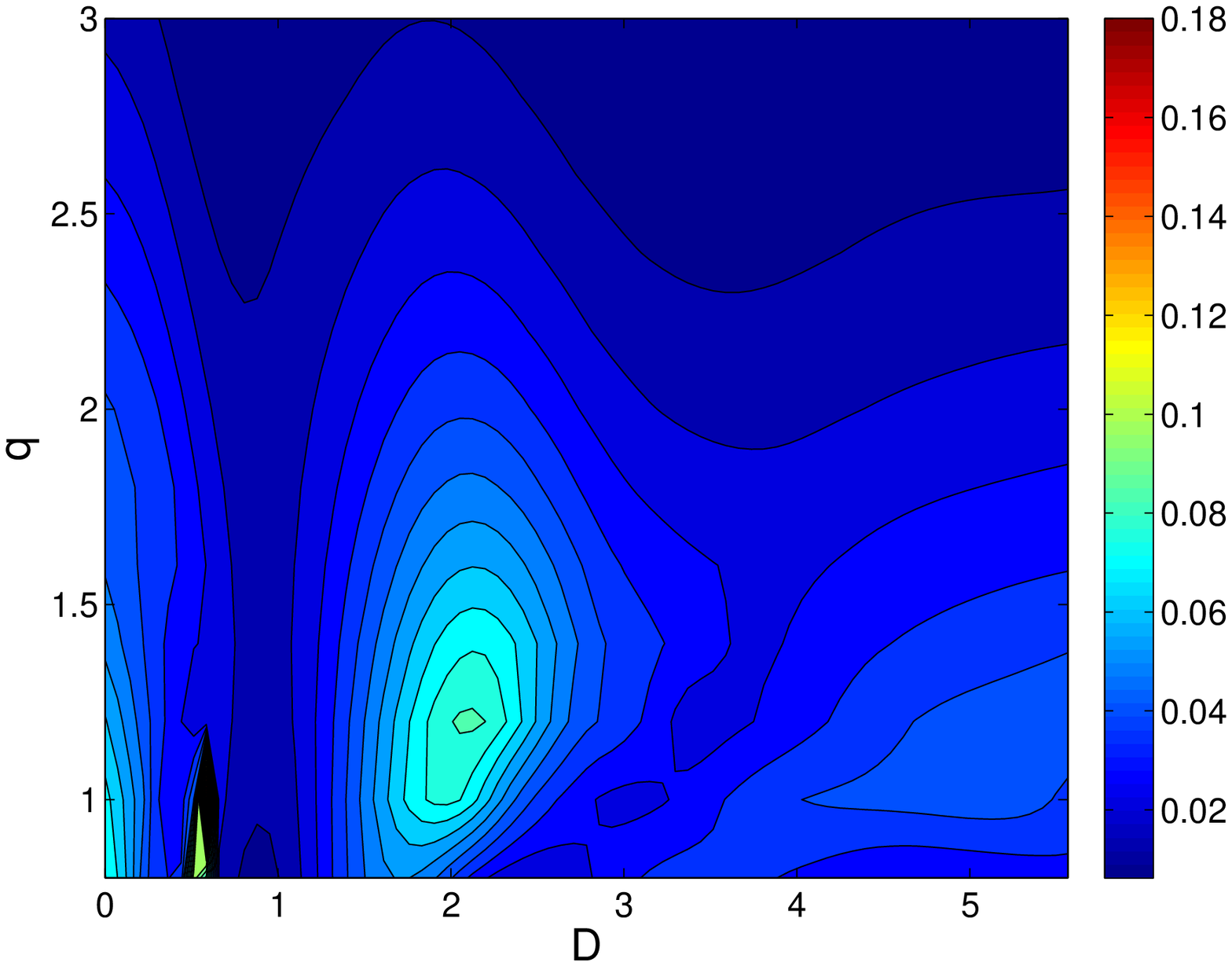} & %
\includegraphics[width=3in]{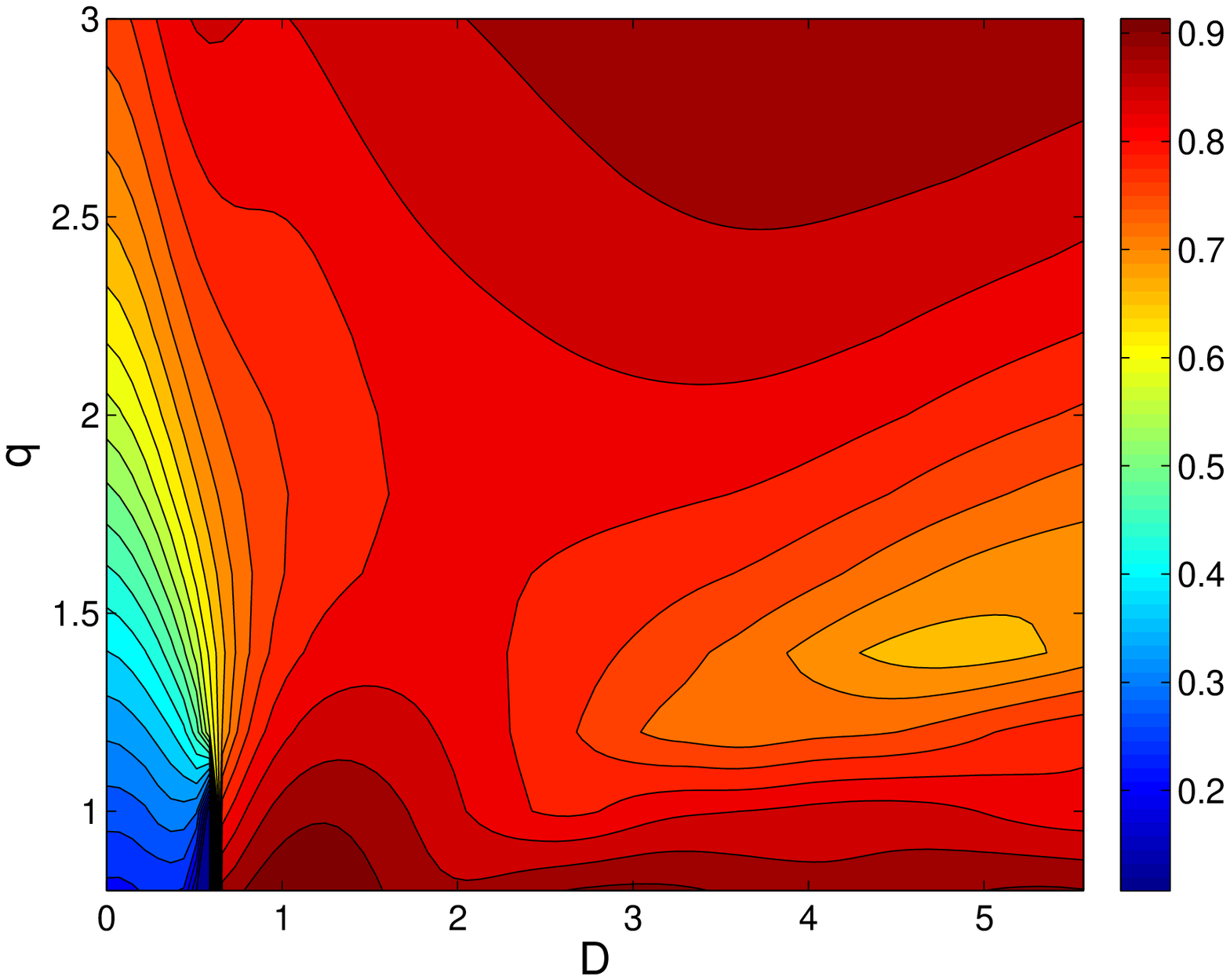} \\
(\mathrm{a}) & (\mathrm{b}) \\
\includegraphics[width=3in]{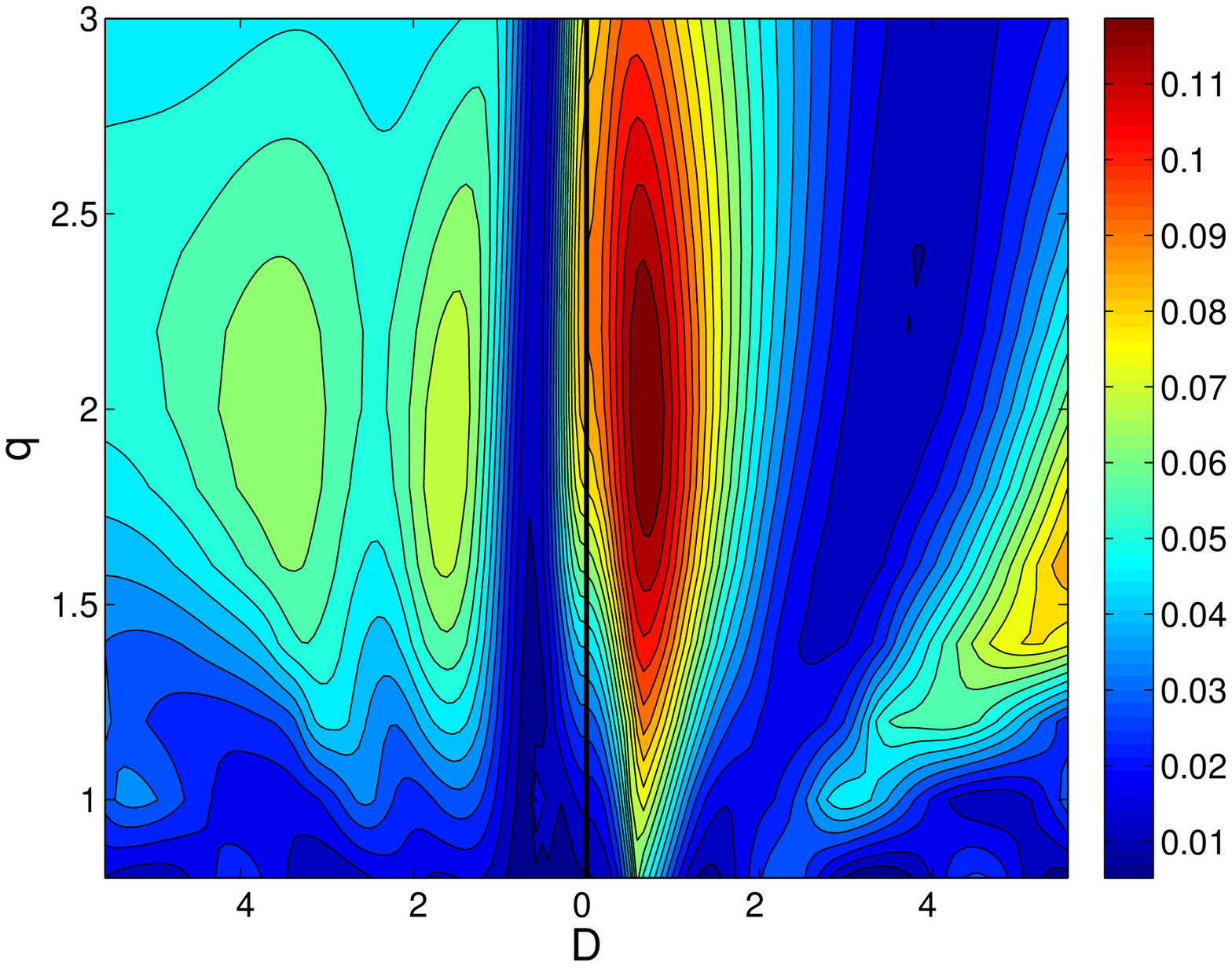}
& \includegraphics[width=3in]{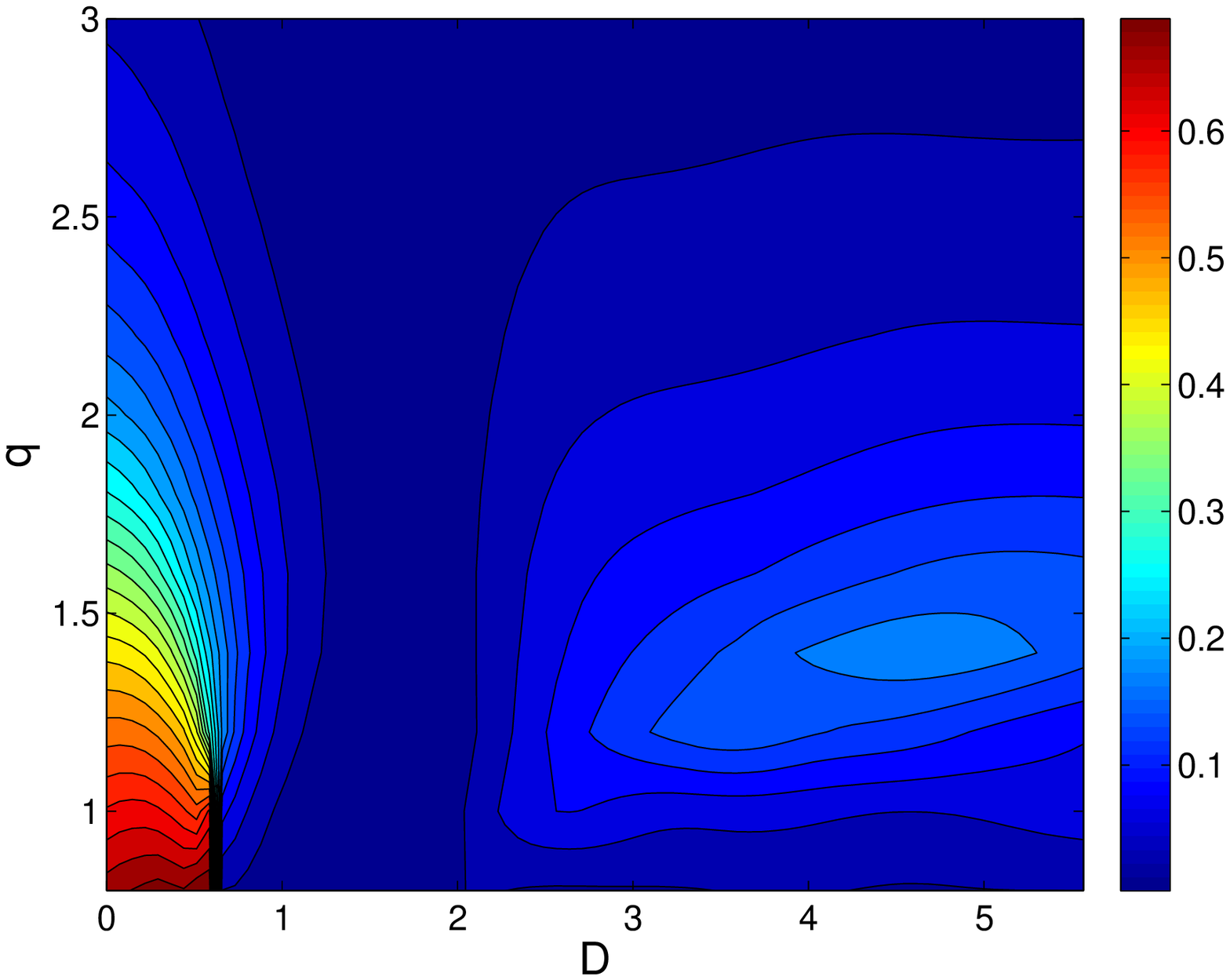}
\\
(\mathrm{c}) & (\mathrm{d})%
\end{array}%
$%
\caption{(Color online) Splitting of the power of the incident soliton, with
amplitude $\protect\eta =2$ and velocity $q$, which hits the pair of local
couplers separated by distance $D$. (a,b) Shares of the power reflected and
transmitted in the straight core. (c) The same for the cross core. In (c),
the plots for the reflected and transmitted shares are butt-joined to
corroborate the equality of these shares for the single coupler ($D=0$). (d)
The trapped-power share.}
\label{FigDualCouplerEta2}
\end{figure}

The increase of the amplitude of the incident soliton to $\eta =2$
significantly reduces the portion of the power transferred to the
cross core, as evident in Fig. \ref{FigDualCouplerEta2} [in
particular, in panel (c)]. In this case, a dominant share of the
incident power is either trapped or transmitted in the straight
core. Accordingly, panels (b) and (d) reveal
the presence a threshold value of the distance between the two couplers, $D_{%
\mathrm{thr}}\approx 0.6$, which is a sharp boundary between the trapping
for $D<D_{\mathrm{thr}}$ and transmission for $D>D_{\mathrm{thr}}$, at $%
q\lesssim 1.2$. Note that such a threshold is not observed at for $\eta =1$,
cf. Fig. \ref{FigDualCoupler}.

If the two couplers are set close enough to each other ($D<0.6$), higher
incident powers give rise to strongly excited trapped states, in which the
power trapped around both couplers swings in unison between the cores. As
the couplers are separated farther, the inter-core-swinging mode changes to
oscillations between the couplers, while the trapped power is greatly
reduced, giving way to a higher power-transmission share in the straight
core.

\begin{figure}[tbp]
%$%
%\begin{array}{cc}
\includegraphics[width=3.5in]{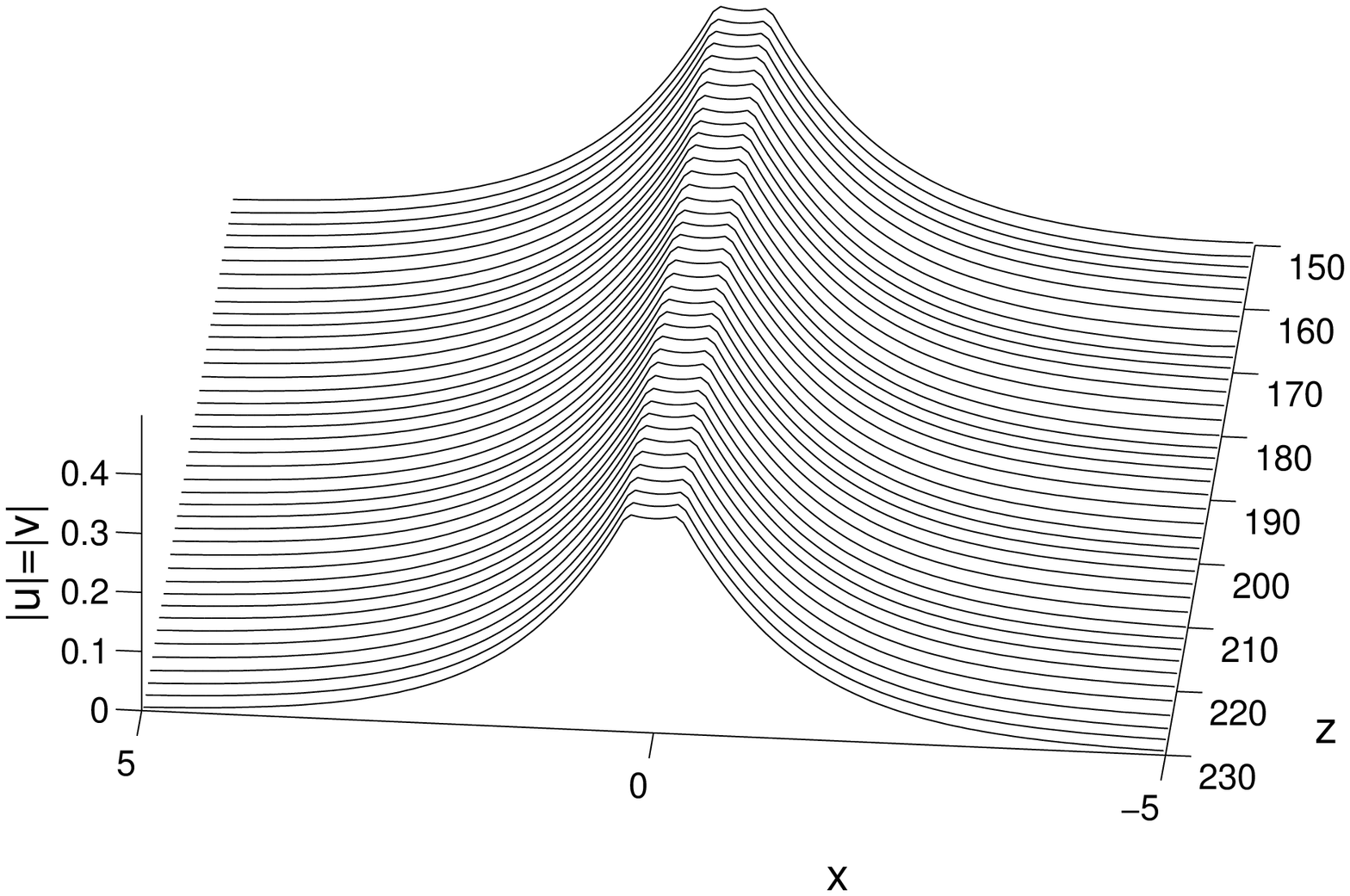}% & %
%\includegraphics[width=3.5in]{scattering_dual_weak.eps} \\
%(\mathrm{a}) & (\mathrm{b})%
%\end{array}%
%$%
\caption{The wave field in the symmetric state, trapped by the dual coupler
of a small width, $D=0.5$, as a result of the passage of a soliton with
amplitude $\protect\eta =1.0$ and initial velocity $q=0.8$.}
\label{FigDualScatter}
\end{figure}

If the distance between the couplers, $D$, is quite small, the field is
trapped in an almost uniform state between them, as shown in Fig.~\ref%
{FigDualScatter}. In this case a solution to the stationary version of Eqs. (%
\ref{u-dual}) and (\ref{v-dual}) outside of the couplers (at $|x|>D/2$) has
the same form as given by Eqs. (\ref{station_exact_form})-(\ref{tantisymm}),
for the symmetric ($U=V$) and asymmetric ($U\neq V$) states alike, while an
exact solution between the couplers, at $|x|<D/2$, can be expressed in terms
of elliptic functions. Actually, for small $D$ the inner solution for the
symmetric state can be approximated by a simple expression which is quadratic in $x$:%
\begin{equation}
U\left( |x|<D/2\right) \approx \sqrt{2k-1}\left\{ 1+\left[ \left( \tfrac{D}{2%
}\right) ^{2}-x^{2}\right] \left( k-1\right) \right\} ,  \label{quadr}
\end{equation}%
and similarly for the asymmetric state. Solution (\ref{quadr}) explains weak
curvature of the intrinsic-layer field observed in Fig.~\ref{FigDualScatter}%
.

It is relevant to mention that the dual coupler may give rise to \textit{%
double symmetry breaking}, combining the SBB of the linearly coupled
components, like in Eqs. (\ref{station_exact_form})-(\ref{tantisymm}), and
spontaneous breaking of the spatial symmetry between the two local couplers,
similar to the well-known symmetry-breaking effect in double-well potentials
\cite{DWP1,Marek}. Analysis of this problem will be presented elsewhere.

It is relevant to mention too that, for $\sqrt{2k}D\gg 1$, the asymmetric
states supported by each local coupler, which are given by Eqs. (\ref%
{station_exact_form}) and (\ref{tantisymm}), can be combined into four
different species of composite states trapped by the pair of far separated
couplers:
\begin{gather}
\left\{ U\left( x\right) ,V\left( x\right) \right\} =\sqrt{2k}\times   \notag
\\
\left\{
\begin{array}{c}
\left\{ \Psi (|x+\tfrac{D}{2}|~+\xi ),\Psi (|x+\tfrac{D}{2}|~+\zeta
)\right\} {\small +}\left\{ \Psi (|x-\tfrac{D}{2}|~+\zeta ),\Psi (|x-\tfrac{D%
}{2}|~+\xi )\right\}  \\
\left\{ \Psi (|x+\tfrac{D}{2}|~+\xi ),\Psi (|x+\tfrac{D}{2}|~+\zeta
)\right\} {\small -}\left\{ \Psi (|x-\tfrac{D}{2}|~+\zeta )],\Psi (|x-\tfrac{%
D}{2}|~+\xi )\right\}  \\
\left\{ \Psi (|x+\tfrac{D}{2}|~+\xi ),\Psi (|x+\tfrac{D}{2}|~+\zeta
)\right\} {\small +}\left\{ \Psi (|x-\tfrac{D}{2}|~+\xi ),\Psi (|x-\tfrac{D}{%
2}|~+\zeta )\right\}  \\
\left\{ \Psi (|x+\tfrac{D}{2}|~+\xi ),\Psi (|x+\tfrac{D}{2}|~+\zeta
)\right\} {\small -}\left\{ \Psi (|x-\tfrac{D}{2}|~+\xi ),\Psi (|x-\tfrac{D}{%
2}|~+\zeta )\right\}
\end{array}%
\right. ,  \label{4}
\end{gather}%
where $\Psi (x)\equiv \mathrm{sech}[\sqrt{2k}(X)]$, while $\xi $ and $\zeta $
are given by Eq. (\ref{tantisymm}) with $+$ and $-$, respectively (hence one
has $0<\xi <\zeta $). The two top lines of Eq. (\ref{4}) represent,
respectively, in-phase and $\pi $-out-of-phase\textit{\ skew composite states%
}, with the opposite sense of the spontaneous symmetry breaking at the two
local couplers: $U(x=-\tfrac{D}{2})>V(x=-\tfrac{D}{2})$ and $\left\vert U(x=%
\tfrac{D}{2})\right\vert <\left\vert V(x=\tfrac{D}{2})\right\vert $ (the $%
\pi $ phase shift implies opposite overall signs of the fields at $x=\pm
\tfrac{D}{2}$), while the two bottom lines represent in-phase and $\pi $%
-out-of-phase\textit{\ straight composite states}, with $U(x=-\tfrac{D}{2}%
)>V(x=-\tfrac{D}{2})$ and $\left\vert U(x=\tfrac{D}{2})\right\vert
>\left\vert V(x=\tfrac{D}{2})\right\vert $. In the inner region, $|x|<\tfrac{%
D}{2}$, decaying tails of the in-phase and out-of-phase states are matched
(like it was done, in another context, in Ref. \cite{Potential}) to the
following solutions of the linearized stationary equations, $k\left\{
U,V\right\} =\tfrac{1}{2}\left\{ U^{\prime \prime },V^{\prime \prime
}\right\} $ [cf. Eq. (\ref{linear_pde_x2})]:%
\begin{eqnarray}
\left\{ U(x),V(x)\right\} _{\mathrm{in}} &=&\left\{ U_{0},V_{0}\right\}
\cosh \left[ \sqrt{2k}\left( x-\left\{ x_{U}^{(0)},x_{V}^{(0)}\right\}
\right) \right] ,  \label{in} \\
\left\{ U(x),V(x)\right\} _{\mathrm{out}} &=&\left\{ U_{0},V_{0}\right\}
\sinh \left[ \sqrt{2k}\left( x-\left\{ x_{U}^{(0)},x_{V}^{(0)}\right\}
\right) \right] .  \label{out}
\end{eqnarray}%
The matching demonstrates that the midpoint of the inner portion of the
straight composite modes is not shifted from the system's center, i.e., $%
x_{U}^{(0)}=x_{V}^{(0)}=0$, while (exponentially small) amplitudes of the
respective inner solutions are%
\begin{equation}
\left\{ U_{0}^{2},V_{0}^{2}\right\} _{\mathrm{straight}}=4\sqrt{2k}\exp %
\left[ -\sqrt{k/2}\left( D+2\left\{ \xi ,\zeta \right\} \right) \right] .
\label{str}
\end{equation}%
On the other hand, for the skew composite states, the midpoint of each
component, $U(x)$ and $V(x)$, is shifted towards the individual soliton with
the smaller amplitude in the same component [i.e., with the amplitude
corresponding to $\zeta $, rather than $\xi $, in Eq. (\ref{tantisymm})]:%
\begin{equation}
x_{U}^{(0)}=-x_{V}^{(0)}=\tfrac{1}{2}\left( \zeta -\xi \right) ,
\end{equation}%
while the corresponding amplitude of the inner solution is
\begin{equation}
\left(
U_{0}^{2}\right) _{\mathrm{skew}}=\left( V_{0}^{2}\right) _{\mathrm{skew}}=4%
\sqrt{2k}\exp \left[ -\sqrt{k/2}\left( D+\xi +\zeta \right)
\right],\nonumber
\end{equation}\
cf. Eq. (\ref{str}).

Finally, following the lines of Ref. \cite{Potential}, it is possible to
compare values of the Hamiltonian of the four species of the composite modes
defined in Eq. (\ref{4}) (the difference in the Hamiltonian is stipulated by
the potential of the interaction between the far separated individually
trapped solitons): $H_{\mathrm{str}}^{\mathrm{(in)}}<H_{\mathrm{skew}}^{%
\mathrm{(in)}}<H_{\mathrm{skew}}^{\mathrm{(out)}}<H_{\mathrm{str}}^{\mathrm{%
(out)}}$, hence the straight in-phase state is the ground state, but the
other species are expected to be dynamically stable too.

\subsection{Caging a shuttle soliton between two local couplers}

The double coupler with large width $D$ can also hold a soliton in the state
of shuttle oscillations, similar to soliton-caging cavities formed by pairs
of far separated barriers, which occur in other models of nonlinear optics
\cite{Chen}. We describe such a cavity by Eqs. (\ref{u-dual}), (\ref{v-dual}%
) without factor $1/2$ in front of $\left[ \delta \left( x-D/2\right)
+\delta \left( x+D/2\right) \right] $, to make each local coupler identical
to the solitary one considered above.

A typical example of the persistent shuttle regime in a broad cavity ($D=40$%
) is shown in Fig.~\ref{FigSingleCoupler}, where a soliton was launched in
the $u$-component as per Eq. (\ref{initialv}), at $\xi _{0}=0$, with $\eta
=0.6$ and $q=0.4$. Note that, with these values of the amplitude and
velocity, the soliton colliding with the local coupler undergoes a nearly
perfect reflection, according to Fig. \ref{FigSingleCoupler}. A dominant
share of the soliton's power stays in the straight core, but some power
penetrates into the cross core, where it forms a small-amplitude wave
pattern whose oscillations are synchronized with the shuttle motion of the
strong component of the soliton. Persistent shuttle regimes are also
possible for symmetric solitons with equal components.

\begin{figure}[tbp]
$%
\begin{array}{cc}
\includegraphics[width=3.5in]{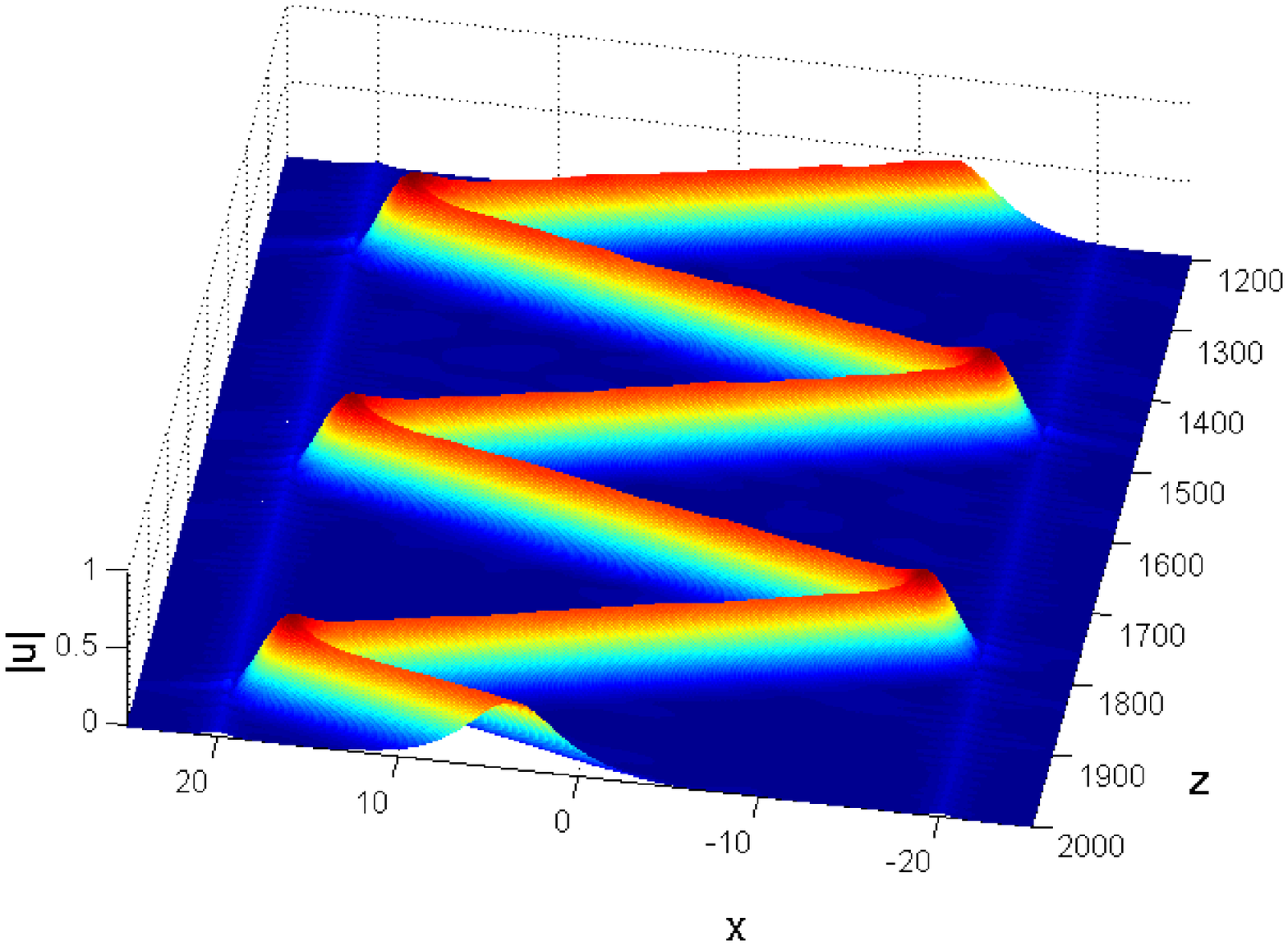} & %
\includegraphics[width=3.5in]{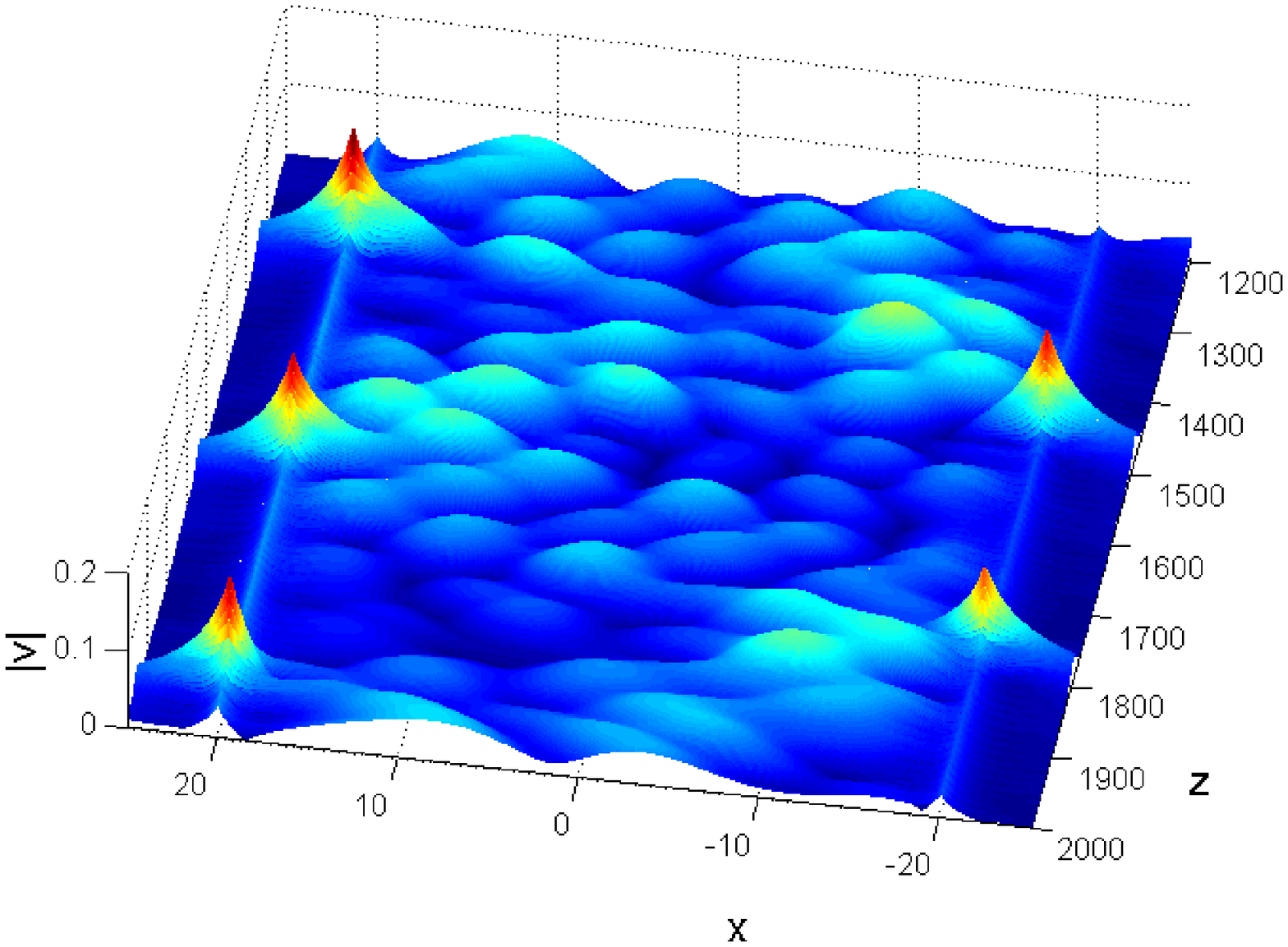} \\
(\mathrm{a}) & (\mathrm{b})%
\end{array}%
$%
\caption{(Color online) An example of persistent shuttle dynamics of a
soliton in a broad cavity ($D=40$) bounded by two local couplers. The
soliton was launched in the $u$-component, with initial amplitude $\protect%
\eta =0.6$ and velocity $q=0.4$. Panels (a) and (b) display absolute
values of fields of $u$ and $v$.} \label{FigCavity}
\end{figure}

\subsection{Shuttle oscillations of solitons in multi-coupler arrays.}

\begin{figure}[tbp]
\includegraphics[width=3.5in]{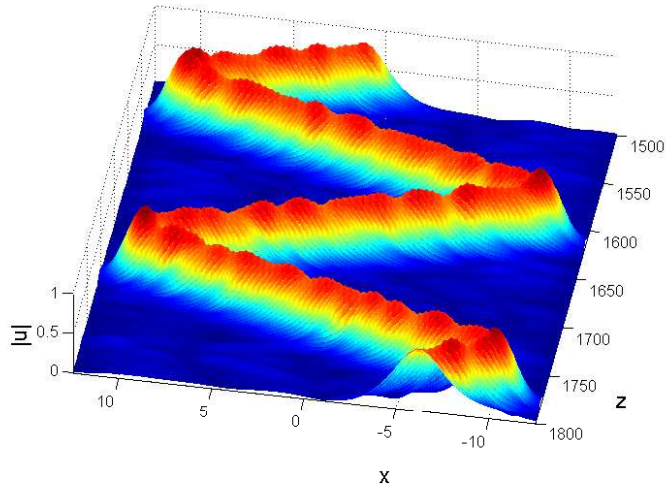}
\caption{(Color online) An example of the persistent shuttle motion of a
symmetric soliton, created with amplitude $\protect\eta =1$ and velocity $%
q=0.8$, in the array defined as per Eq. (\protect\ref{C}), with $C=1/4$ and $%
D=0.354$. The number of local couplers in the array is $2N+1=65$. Note that
these parameters satisfy the average trapping condition (\protect\ref{box}).}
\label{FigCouplerArrayScattering}
\end{figure}

An example displayed in Fig. \ref{FigCouplerArrayScattering} demonstrates
that solitons can also be held in the state of persistent shuttle motion in
the array defined as per Eq. (\ref{C}), periodically bouncing from edges of
the array. In the example shown in this figure, the soliton is symmetric,
hence the potential of its interaction with a solitary local coupler is $W_{%
\mathrm{coupl}}^{\mathrm{(symm)}}(\xi )=-2C\eta ^{2}\mathrm{sech}^{2}\left(
\eta \xi \right) $, see Eq. (\ref{attraction}). For a broad soliton, with $%
\eta ^{-1}\gg D$, the potentials induced by the comb of couplers can be
averaged, making the array tantamount to a potential box of depth $4C\eta /D$
and spatial width $2ND$, see Eq. (\ref{Xbox}). Thus, the symmetric soliton
with the double effective mass, $M^{\mathrm{(symm)}}=4\eta $, is trapped in
the box if its squared velocity takes values
\begin{equation}
q^{2}<2C/D.  \label{box}
\end{equation}%
Note that this trapping threshold does not depend on the soliton's
amplitude, $\eta $.

For strongly asymmetric solitons, with mass $M\approx 2\eta $, the effective
potential of the interaction with the solitary coupler is given by Eq. (\ref%
{adia_full_inter}), which should be here multiplied by $C^{2}$. Thus, the
average depth of the respective potential box is $3C^{2}/D$, and trapping
condition (\ref{box}) is replaced by $q^{2}<3C^{2}/\left( D\eta \right) $,
which this time depends on $\eta $.

\section{Conclusion}

The objective of this work is to extend the well-elaborated analysis of the
dynamics of solitons in nonlinear couplers to dual-core waveguides fused at
one or several narrow segments, which can be approximated by $\delta $%
-functions. The model directly applies to optical spatial solitons in
dual-core planar waveguides with the Kerr nonlinearity, as well as to dual
traps for the self-attracting BEC. By means of systematic simulations and a
combination of several analytical approximations, we have studied collisions
of the free soliton with single and double local couplers. The outcome of
the collision is characterized by splitting of the total power between five
waves, \textit{viz}., those transmitted and reflected in each core of the
waveguide, and the one trapped by the local waveguide. Dynamics of the
soliton trapped by the local coupler was studied too, by means of the
variational approximation and simulations. Shuttle oscillations of a soliton
between two local couplers, and the shuttle motion between edges of a finite
array of couplers, were also addressed. In the context of the pair of far
separated local couplers, four species of straight and skew trapped
two-soliton states were predicted analytically.

The present analysis can be extended in other directions. In particular, it
may be interesting, as mentioned above, to study the double symmetry
breaking in the double local coupler. On the other hand, it may be also
relevant to introduce a similar model with the quadratic nonlinearity.

\end{document}